\documentclass[aps,prb,superscriptaddress,showpacs,floatfix,twocolumn,amsmath]{revtex4-1}
\usepackage[utf8]{inputenc}

\usepackage{amsmath}
\usepackage{physics}
\usepackage{braket}
\usepackage{graphicx}
\usepackage{color}
\usepackage{siunitx}
\usepackage{dsfont}
\usepackage{multirow}
\usepackage{mathtools}

\newcommand{\etal}{\textit{et al}}

\newcommand{\eul}{\text{e}}
\newcommand{\ci}{\text{i}}
\newcommand{\id}{\mathds{1}}
\newcommand{\nk}{{n \vb k}}
\newcommand{\plka}{{\partial_{l \kappa \alpha}}}

\newcommand{\bzint}{{\int_\text{BZ}}}
\newcommand{\BZvol}{{\Omega_\text{BZ}}}
\newcommand{\kint}[1]{{\frac{\dd[3]{#1}}{\BZvol}}}
\newcommand{\diracv}[1]{{\delta^\qty(3)\qty(#1)}}
\newcommand{\paw}[1]{\widetilde{#1}}
\newcommand{\pawT}{\hat{\mathcal{T}}}
\newcommand{\pPsi}{\paw{\Psi}}

\begin{document}

\title{Electron-Phonon Interactions Using the PAW Method and Wannier Functions}

\author{Manuel Engel}
\author{Martijn Marsman}
\author{Cesare Franchini}
\author{Georg Kresse}
\affiliation{University of Vienna, Faculty of Physics and Center for Computational Materials Science, Sensengasse 8, A-1090 Vienna, Austria}

\begin{abstract}
	We present an ab-initio density-functional-theory approach for calculating electron-phonon interactions within the projector augmented-wave method.
	The required electron-phonon matrix elements are defined as the second derivative of the one-electron energies in the PAW method.
	As the PAW method leads to a generalized eigenvalue problem, the resulting electron-phonon matrix elements lack some symmetries that are usually present for simple eigenvalue problems and all-electron formulations.
	We discuss the relation between our definition of the electron-phonon matrix element and other formulations.
	To allow for efficient evaluation of physical properties, we introduce a Wannier-interpolation scheme, again adapted to generalized eigenvalue problems.
	To explore the method's numerical characteristics, the temperature-dependent band-gap renormalization of diamond is calculated and compared with previous publications.
	Furthermore, we apply the method to selected binary compounds and show that the obtained zero-point renormalizations agree well with other values found in literature and experiments.
\end{abstract}

\maketitle

\section{Introduction}

Electron-phonon interactions are exceedingly relevant for the simulation of both zero and finite-temperature properties~\cite{giustino-RevModPhys, grimvall_elphon, vogl_elphon}.
Be it their theoretical importance in understanding fundamental aspects of condensed matter or their great practical value in developing novel and superior technologies, electron-phonon interactions are indispensable in modern physics.
They provide a rigorous framework for describing thermal transport in semiconductors, the electrical resistivity's temperature dependence in metals, explain the emergence of conventional superconductivity and are responsible for the formation of polarons, to name a few examples.
Furthermore, the importance of electron-phonon interactions in developing batteries~\cite{batteries-1,batteries-2}, solar cells~\cite{solar-cells-1,solar-cells-2}, thermoelectrics~\cite{thermoelectrics-1,thermoelectrics-2,thermoelectrics-3,thermoelectrics-4} and organic electronics~\cite{organic-electronics} highlights their practical significance.

While their theoretical importance is unquestionable, the inclusion of electron-phonon interactions in routine ab-initio simulations has been a relatively recent development.
A lack of efficient ab-initio algorithms and the wide-spread use of semi-empirical model Hamiltonians were part of the reason for this delay in development.
Nevertheless nowadays, there exists a broad spectrum of available computational models that promise to describe electron-phonon interactions.
They can roughly be divided into two categories, as the methods are either based on statistical sampling or perturbation theory.

For sampling methods, the conceptually simplest approach is to average over atomic configurations from statistical ensembles.
This can be achieved in a multitude of ways, for example, via Monte-Carlo (MC) integration~\cite{MC-1, MC-2} or molecular dynamics (MD)~\cite{MD-1, MD-2}.
Methods based on statistical sampling are usually easy to implement and are capable of capturing anharmonic effects~\cite{anharmonic}.
Due to their statistical nature, however, they also feature substantial disadvantages.
In particular, their reliance on large supercells and, potentially, vast numbers of samples often incurs subpar computational scaling with regard to system size.
Furthermore, using a sampling method might not be appropriate when the quantity of interest needs to be resolved with respect to individual phonon modes.
The need for many samples has recently been challenged by a novel one-shot method~\cite{giustino-one-shot} that only requires a single supercell configuration.
It allows for efficient computations of the phonon-induced renormalization of ab-initio electronic band structures if long-range electrostatic effects are not relevant (e.g.~Fröhlich-like interactions).

Alternatively, electron-phonon interactions can be calculated using perturbation theory.
Contrary to the simplicity offered by sampling methods, approaches based on perturbation theory are usually more involved.
In most cases, however, they are substantially faster and provide direct computational access to intermediate quantities.
The de-facto standard for treating electron-phonon interactions in this context is density-functional perturbation theory (DFPT)~\cite{DFPT, gonze_DFPT}.
This may involve merely a single unit cell and allows for the calculation of phonon-related quantities at arbitrary wave vectors.

Many crystal-related properties, such as finite-temperature transport, heat conductivity and superconducting properties require a very fine sampling of the Brillouin zone.
This can significantly increase the computational cost if methods such as DFPT are used directly for a large number of phonon wave vectors.
Fortunately, there exist interpolation methods based on Wannier functions~\cite{giustino-coupling,elphon-wannier-interpol,wannier-functions,mauri_interpol} or atomic orbitals~\cite{atomic_orbital_interpol} that offer a tremendous decrease in computation time while maintaining reasonable accuracy.
Starting from only a small number of potentially expensive ab-initio steps, this method allows the calculation of electron-phonon matrix elements at arbitrary wave vectors through interpolation.

We aspire to build an implementation based on perturbation theory and Wannier orbitals, distinguished from others by the employment of the projector augmented-wave (PAW) method~\cite{bloechl, kresse-PAW} and a finite-difference scheme in real space.
Despite having to rely on using large super cells, this approach has multiple advantages.
The PAW method strikes an excellent compromise between speed and precision, rivaling the accuracy of FLAPW methods~\cite{FLAPW}.
Furthermore, finite differences are universally applicable to any functional, including hybrid or meta-gradient functionals, since linear-response theory is avoided.
Finally, the implementation is integrated into the Vienna Ab-initio Simulation Package (VASP)~\cite{VASP-1,VASP-2,VASP-3,VASP-4}, which boasts a solid and feature-rich simulation environment.

The approach is tested for a number of materials.
We seize this opportunity to compare numerical results with the one-shot method previously implemented in VASP~\cite{one-shot-Karsai}.
This provides both a way to verify the credibility of our implementation as well as a meaningful way of comparing the two computational approaches as they are part of the same software package.

The outline of the paper is as follows.
In Sec.~\ref{sec:electron_self_energy}, the phonon-induced renormalization of the electronic band structure is derived within the PAW framework using perturbation theory.
This leads to the definition of a PAW electron-phonon matrix element that is discussed in more detail in Sec.~\ref{sec:matrix_element}.
Subsequently, the Wannier-interpolation algorithm used to calculate such matrix elements is outlined in Sec.~\ref{eq:wannier_interpolation}.
Numerical results and a discussion thereof are presented in Sec.~\ref{sec:results}, followed by a short conclusion in Sec.~\ref{sec:conclusion}.

\section{Electron Self Energy in the PAW Framework}\label{sec:electron_self_energy}

Most ab-initio methods for determining the electronic ground state rely on the Born--Oppenheimer approximation that completely neglects the ionic vibrational degrees of freedom.
A common approach aiming to include lattice dynamics and electron-phonon interactions is to expand the quantity of interest in a perturbation series.
The derivation presented here follows the spirit of the perturbation theory developed by Allen, Heine and Cardona (AHC)~\cite{allen-heine, allen-cardona}.

To begin with, the fully temperature-dependent energy, \(\varepsilon_\nk \qty(T) \), of a single Kohn--Sham electron with band index \(n \) and Bloch vector \(\vb k \) is expanded with respect to individual atomic displacements, \(\vb u_{l\kappa} \),
\begin{equation}
	\label{eq:energy_expansion}
	\begin{split}
		\varepsilon_\nk\qty(T)
		& =
		\varepsilon_\nk
		+
		\sum_{l \kappa \alpha} \pdv{\varepsilon_\nk}{u_{l\kappa\alpha}} \ev{u_{l \kappa \alpha}}_T
		\\ & +
		\frac{1}{2}
		\sum_{l \kappa \alpha} \sum_{l' \kappa' \beta}
		\pdv{\varepsilon_\nk}{u_{l\kappa\alpha}}{u_{l'\kappa'\beta}}
		\ev{u_{l \kappa \alpha} u_{l' \kappa' \beta}}_T
		\\ & + \ldots
		\, ,
	\end{split}
\end{equation}
where \(\varepsilon_\nk \) are the Kohn--Sham (KS) eigenvalues of the purely electronic ground state and \(l \), \(\kappa \), \(\alpha \) are cell, atom and Cartesian indices, respectively.
Where appropriate, these indices are absorbed into a compound index, \(\tau \equiv \qty(l, \kappa, \alpha) \), and the shorthand notation \(\partial_\tau \equiv \pdv{}{u_{l\kappa\alpha}} \) is used.
By convention, \(\partial_\tau \) here only acts on the term immediately following the differential operators.

Neglecting the first-order, third-order as well as all higher-order terms in Eq.~\eqref{eq:energy_expansion} leaves the quadratic, harmonic term as the sole contribution to the shift in energy, \(\Delta \varepsilon_\nk \qty(T) \), due to electron-phonon interactions:
\begin{equation}
	\label{eq:energy_shift}
	\Delta \varepsilon_\nk\qty(T)
	\equiv
	\frac{1}{2}
	\sum_{\tau} \sum_{\tau'} \partial_\tau \partial_{\tau'} \varepsilon_\nk \ev{u_{\tau} u_{\tau'}}_T
	\, .
\end{equation}
Needless to say, this approximation breaks down if the system exhibits strong anharmonicities.
However, this formulation allows for a simple treatment of the vibrational degrees of freedom in terms of phonons.

To begin with, we focus on the thermal expectation value \(\ev{u_{\tau} u_{\tau'}}_T \).
Using phonon creation and annihilation operators, it is straightforward to cast this term into a useful expression,
\begin{equation}
	\label{eq:thermal_expect}
	\begin{split}
		\ev{u_{l \kappa \alpha} u_{l' \kappa' \beta}}_T
		& =
		\bzint \kint{q} \sum_\nu
		\frac{\hbar \qty(n_{\nu \vb q}\qty(T) + \frac{1}{2})}{\omega_{\nu \vb q} \sqrt{m_\kappa m_\kappa'}}
		\\ & \cross
		\eul^{\ci \vb q \vdot \qty(\vb R_{l \kappa} - \vb R_{l' \kappa'})} 
		e_{\kappa \alpha, \nu \vb q} e^*_{\kappa' \beta, \nu \vb q}
		\, ,
	\end{split}
\end{equation}
with \(\omega_{\nu \vb q} \) and \(e_{\kappa\alpha, \nu \vb q} \) being the angular frequencies and eigenvectors, respectively, of a phonon with wave vector \(\vb q \) and branch index \(\nu \).
\(m_\kappa \) is the mass of ion \(\kappa \), \(\vb R_{l\kappa} \) the equilibrium position of that ion in cell \(l \) and \(n_{\nu \vb q}\qty(T) \) is the Bose--Einstein distribution function for a phonon with energy \(\hbar \omega_{\nu \vb q} \).

Equation~\eqref{eq:thermal_expect} describes the average thermal fluctuations of the ionic system in terms of independent phonon modes.
It is the only temperature-dependent quantity that enters the energy shift in Eq.~\eqref{eq:energy_shift}.
Obviously, for \(T \to 0 \Rightarrow n_{\nu \vb q}\qty(T) \to 0 \), there is still a contribution due to zero-point vibrations of the lattice, commonly referred to as zero-point renormalization (ZPR).

Recalling Eq.~\eqref{eq:energy_shift}, the second derivative of the KS eigenvalues is yet to be determined.
In the PAW framework, the KS equations generalize to
\begin{equation}
	\label{eq:PAW_KS_equation}
	\paw{H} \ket{\pPsi_\nk} = \varepsilon_\nk \paw{S} \ket{\pPsi_\nk}
	\, ,
\end{equation}
where \(\paw{H} \) and \(\paw{S} \) are the PAW Hamiltonian and overlap operator, respectively, and \(\ket{\paw{\Psi}_\nk} \) are smooth pseudo Bloch orbitals.
These quantities are related to their all-electron (AE) counterparts via the usual linear transformation, \(\pawT \):
\begin{align}
	\label{eq:PAW_transform}
	\ket{\Psi_\nk} & = \pawT \ket{\pPsi_\nk} \\
	\paw{H} & = \pawT^\dagger \hat{H} \pawT \\
	\paw{S} & = \pawT^\dagger \pawT
	\, .
\end{align}

Differentiation of Eq.~\eqref{eq:PAW_KS_equation} with respect to a single atomic displacement, \(u_{l\kappa\alpha} \), yields a special instance of the Hellman--Feynman theorem for the PAW method:
\begin{equation} \label{eq:hellmann_feynman}
	\partial_\tau \varepsilon_\nk =
	\braket{\pPsi_\nk | \partial_\tau \paw{H} - \varepsilon_\nk \partial_\tau \paw{S} | \pPsi_\nk}
	\, .
\end{equation}
Repeating this process once again leads to an expression for the second derivative that can be split into two distinct contributions,
\begin{equation}
	\label{eq:FM_DW_contribution_split}
	\partial_\tau \partial_{\tau'} \varepsilon_\nk =
	\delta \varepsilon^\text{FM}_{\nk, \tau \tau'} +
	\delta \varepsilon^\text{DW}_{\nk, \tau \tau'}
	\, ,
\end{equation}
where \(\delta \varepsilon^\text{FM}_{\nk, \tau \tau'} \) and \(\delta \varepsilon^\text{DW}_{\nk, \tau \tau'} \) will be referred to as the Fan--Migdal (FM) and Debye--Waller (DW) contributions, respectively.
The corresponding energy shifts, \(\Delta \varepsilon^\text{FM}_\nk \) and \(\Delta \varepsilon^\text{DW}_\nk \), are given by Eq.~\eqref{eq:energy_shift}:
\begin{align}
	\Delta \varepsilon_\nk\qty(T) & \equiv \Delta \varepsilon^\text{FM}_\nk\qty(T) + \Delta \varepsilon^\text{DW}_\nk\qty(T)
	\, , \\ \label{eq:FM_energy_shift}
	\Delta \varepsilon^\text{FM}_\nk\qty(T) & \equiv
	\frac{1}{2} \sum_{\tau} \sum_{\tau'} \delta\varepsilon^\text{FM}_{\nk, \tau \tau'}
	\ev{u_{\tau} u_{\tau'}}_T
	\, , \\ \label{eq:DW_energy_shift}
	\Delta \varepsilon^\text{DW}_\nk\qty(T) & \equiv
	\frac{1}{2} \sum_{\tau} \sum_{\tau'} \delta\varepsilon^\text{DW}_{\nk, \tau \tau'}
	\ev{u_{\tau} u_{\tau'}}_T
	\, .
\end{align}

The FM contribution subsumes all terms containing factorizable first derivatives,
\begin{multline}
	\label{eq:FM_contribution}
	\delta \varepsilon^\text{FM}_{\nk, \tau \tau'}
	\equiv
	-
	\paw{h}_{\nk, \nk, \tau} \paw{h}^\text{S}_{\nk, \nk, \tau'}
	-
	\paw{h}_{\nk, \nk, \tau'} \paw{h}^\text{S}_{\nk, \nk, \tau}
	\\ +
	\sum'_{m\vb k'}
	\frac{
		\paw{h}^*_{m \vb k' \nk, \tau}
		\paw{h}_{m \vb k' \nk, \tau'}
	}
	{\varepsilon_{\nk} - \varepsilon_{m \vb k'}}
	+
	\sum'_{m\vb k'}
	\frac{
		\paw{h}^*_{m \vb k' \nk, \tau'}
		\paw{h}_{m \vb k' \nk, \tau}
	}
	{\varepsilon_{\nk} - \varepsilon_{m \vb k'}}
	\, ,
\end{multline}
with
\begin{align}
	\label{eq:shorthand_h}
	\paw{h}_{m \vb k' \nk, \tau}
	& \equiv
	\braket{\pPsi_{m \vb k'} | \partial_\tau \paw{H} - \varepsilon_\nk \partial_\tau \paw{S} | \pPsi_\nk}
	\, ,
	\\
	\label{eq:shorthand_hS}
	\paw{h}^\text{S}_{m \vb k' \nk, \tau}
	& \equiv
	\braket{\pPsi_{m \vb k'} | \partial_\tau \paw{S} | \pPsi_\nk}
	\, .
\end{align}
A detailed derivation is given in Appendix~\ref{sec:appendix:FM_DW_contributions}.
The primed sum appearing in Eq.~\eqref{eq:FM_contribution} excludes all divergent terms with \(\qty(\nk) = \qty(m \vb k') \) and the domain of integration spans the first Brillouin zone with volume \(\BZvol \).
In practice, potential divergence problems due to degeneracies can be avoided by introducing a small imaginary shift in the denominator.
Finally, the remaining term containing only second derivatives comprises the DW contribution:
\begin{equation}
	\label{eq:DW_contribution}
	\delta \varepsilon^\text{DW}_{\nk, \tau \tau'}
	\equiv
	\braket{\pPsi_\nk | \partial_\tau \partial_{\tau'} \paw{H} - \varepsilon_\nk \partial_\tau \partial_{\tau'} \paw{S} | \pPsi_\nk}
	\, .
\end{equation}

It is worth mentioning that the aforementioned FM and DW contributions are, in fact, not identical to the ones usually obtained from non-generalized eigenvalue problems.
Nevertheless, the presented formulation guarantees numerical stability for calculating derivatives inside the PAW framework.
In Sec.~\ref{sec:matrix_element}, this aspect is briefly revisited.

As expected, the FM as well as the DW contribution are manifestly invariant under the exchange of the compound indices \(\tau \) and \(\tau' \).
Additionally, it is easy to confirm that by setting \(\paw{S} \) equal to the identity operator, one obtains the usual results for norm-conserving pseudopotentials.
In this case, the PAW transformation operators are necessarily unitary and all terms of the form \(\partial_\tau \paw{S} \) vanish.

At this point, it would be desirable to express the FM and DW contributions in terms of individual phonon modes, similar to Eq.~\eqref{eq:thermal_expect}.
To this end, one introduces the differential operator
\begin{equation}
	\label{eq:derivative_phonons}
	\partial_{\nu \vb q}
	\equiv
	\sqrt{\frac{\hbar}{2 \omega_{\nu \vb q}}}
	\sum_{l\kappa\alpha} \frac{1}{\sqrt{m_\kappa}}
	e_{\kappa\alpha, \nu \vb q} \eul^{\ci \vb q \vdot \vb R_{l\kappa}}
	\plka
	\, ,
\end{equation}
which corresponds to a collective ionic displacement along the phonon mode with wave vector \(\vb q \) and branch index \(\nu \).
We will refer to it as the phonon displacement operator.
Equation~\eqref{eq:derivative_phonons} is easily inverted to yield an expression for the ionic displacement operator, \(\plka \), in terms of the phonon displacement operator:
\begin{equation} \label{eq:derivative_ions}
	\plka =
	\sqrt{m_\kappa} \bzint \kint{q} \sum_{\nu}
	\sqrt{\frac{2 \omega_{\nu \vb q}}{\hbar}}
	e^*_{\kappa\alpha, \nu \vb q} \eul^{-\ci \vb q \vdot \vb R_{l\kappa}}
	\partial_{\nu \vb q}
	\, .
\end{equation}
Finally, using Eq.~\eqref{eq:FM_energy_shift},~\eqref{eq:FM_contribution} and~\eqref{eq:derivative_ions}, the FM self energy can be written as
\begin{multline}
	\label{eq:FM_self_energy}
	\Delta \varepsilon^\text{FM}_\nk\qty(T)
	=
	- \sum_\nu
	\paw{g}_{nn \vb k, \nu \vb 0}
	\paw{g}^\text{S}_{nn \vb k, \nu \vb 0}
	\qty(2 n_{\nu \vb 0}\qty(T) + 1)
	\\ +
	\bzint \kint{q} \sum_\nu  \sum_{m}'
	\frac{\abs{\paw g_{mn \vb k, \nu \vb q}}^2}{\varepsilon_\nk - \varepsilon_{m \vb k + \vb q}}
	\qty(2 n_{\nu \vb q}\qty(T) + 1)
	\, ,
\end{multline}
with
\begin{align}
	\label{eq:g_PAW}
	\paw{g}_{mn \vb k, \nu \vb q}
	& \equiv
	\braket{\pPsi_{m \vb k + \vb q} | \partial_{\nu \vb q} \paw H - \varepsilon_{\nk} \partial_{\nu \vb q} \paw S | \pPsi_\nk}
	\\
	\label{eq:gS_PAW}
	\paw{g}^\text{S}_{mn \vb k, \nu \vb q}
	& \equiv
	\braket{\pPsi_{m \vb k + \vb q} | \partial_{\nu \vb q} \paw S | \pPsi_\nk}
	\, .
\end{align}
The quantity \(\paw{g}_{mn\vb k, \nu\vb q} \) will be referred to as PAW electron-phonon matrix element.
Its properties are discussed in more detail in Sec.~\ref{sec:matrix_element}.

Retrieving the DW self energy in a similar fashion is more involved, since Eq.~\eqref{eq:DW_energy_shift} involves a double sum over all atoms of the lattice.
In practice, the amount of independent displacements renders this expression challenging to compute efficiently such that approximations become necessary.

Allen and Heine~\cite{allen-heine} have used the invariance of the total energy with respect to lattice translations to relate the FM to the DW contribution.
A generalization to crystals featuring a multi-atom basis was subsequently suggested by Allen and Cardona~\cite{allen-cardona}.
Allen, Heine and Cardona used the rigid-ion approximation to retrieve a workable expression for the DW self energy.
The crucial simplification is that the second derivative, Eq.~\eqref{eq:DW_contribution}, shall only yield on-site contributions, e.g.
\begin{equation}
	\delta \varepsilon^\text{DW}_{n\vb k, l\kappa\alpha l'\kappa'\beta} = 0
	\qq{if}
	l \neq l' \lor \kappa \neq \kappa'
	\, .
\end{equation}
The DW contribution can thus be calculated similarly to the FM contribution.
Carrying out the algebra results in
\begin{multline}
	\label{eq:DW_self_energy}
	\Delta \varepsilon^\text{DW}_\nk\qty(T)
	=
	\bzint \kint{q} \sum_\nu
	\Xi^\text{S}_{n\vb k, \nu \vb q}
	\qty(2 n_{\nu \vb q}\qty(T) + 1)
	\\ -
	\bzint \kint{q} \sum_\nu  \sum_{m}'
	\frac{\Xi_{mn\vb k, \nu \vb q}}{\varepsilon_\nk - \varepsilon_{m \vb k}}
	\qty(2 n_{\nu \vb q}\qty(T) + 1)
	\, ,
\end{multline}
where
\begin{align}
	\Xi_{mn\vb k, \nu \vb q}
	& \equiv
	\frac{\hbar}{4 \omega_{\nu\vb q}}
	\sum_{\kappa\alpha} \sum_{\kappa' \beta}
	\Theta_{\kappa\alpha, \kappa'\beta}^{\nu \vb q}
	\paw{g}^{0*}_{mn\vb k, \kappa \alpha}
	\paw{g}^0_{mn\vb k, \kappa' \beta}
	\, , \\
	\Xi^\text{S}_{n\vb k, \nu \vb q}
	& \equiv
	\frac{\hbar}{4 \omega_{\nu\vb q}}
	\sum_{\kappa\alpha} \sum_{\kappa' \beta}
	\Theta_{\kappa\alpha, \kappa'\beta}^{\nu \vb q}
	\paw{g}^0_{nn\vb k, \kappa \alpha}
	\paw{g}^{\text{S}0}_{nn\vb k, \kappa' \beta}
	\, , \\
	\label{eq:DW_Theta_definition}
	\Theta_{\kappa\alpha, \kappa'\beta}^{\nu \vb q}
	& \equiv
	\frac{e_{\kappa\alpha, \nu\vb q} e^*_{\kappa\beta, \nu\vb q}}{m_\kappa}
	+
	\frac{e^*_{\kappa'\alpha, \nu\vb q} e_{\kappa'\beta, \nu\vb q}}{m_{\kappa'}}
	\, , \\
	\label{eq:DW_g0_definition}
	\paw{g}^0_{mn\vb k, \kappa \alpha}
	& \equiv
	\sum_\nu
	\sqrt{\frac{2 m_\kappa \omega_{\nu\vb 0}}{\hbar}}
	e_{\kappa \alpha, \nu\vb 0} \paw g_{mn\vb k, \nu\vb 0}
	\, , \\
	\label{eq:DW_gS0_definition}
	\paw{g}^{\text{S}0}_{mn\vb k, \kappa \alpha}
	& \equiv
	\sum_\nu
	\sqrt{\frac{2 m_\kappa \omega_{\nu\vb 0}}{\hbar}}
	e_{\kappa \alpha, \nu\vb 0} \paw g^\text{S}_{mn\vb k, \nu\vb 0}
	\, .
\end{align}
Equation~\eqref{eq:DW_self_energy} has been written such that it takes a form reminiscent of Eq.~\eqref{eq:FM_self_energy}.
As in the case of the FM self energy, additional terms appear due to the PAW method.
These terms, while obviously different, still retain some similarity to the ones appearing in the FM self energy.
A more detailed derivation is found in appendix~\ref{sec:appendix:self_energy}.

\section{Electron-Phonon Matrix Element in the PAW Framework}\label{sec:matrix_element}

In Sec.~\ref{sec:electron_self_energy}, the electron-phonon matrix element has been defined via the second derivative of the one-electron energies with respect to the phonon displacement operator.
In this section, a closer look is taken at some of its more peculiar properties and the nuances arising from the PAW formalism.

In the context of DFT, a common definition for the electron-phonon matrix element is~\cite{giustino-RevModPhys}
\begin{equation} \label{eq:ae_epme}
	g_{mn\vb k,\nu\vb q} \equiv \braket{\Psi_{m \vb k + \vb q} | \partial_{\nu\vb q} \hat H | \Psi_\nk}
	\, .
\end{equation}
Chaput, Togo and Tanaka recently suggested to use the following formally equivalent matrix element\cite{elphon_PAW_chaput}
\begin{equation} \label{eq:ae_epme}
 g_{mn\vb k,\nu\vb q} = \qty(\varepsilon_\nk - \varepsilon_{m\vb k + \vb q}) \braket{\Psi_{m \vb k + \vb q} | \partial_{\nu\vb q} \Psi_\nk}
	\, ,
\end{equation}
which will be referred to as AE matrix element.
They suggested to insert the PAW transformation \(\Psi_\nk = \pawT \ket{\pPsi_\nk} \) in this equation and then use the PAW completeness relation to simplify the matrix element.
This approach implies that explicit derivatives of the operator \(\pawT \), and thus of the partial waves, need to be calculated.

Our \(\paw{g}_{mn\vb k,\nu\vb q} \) are formally equivalent to 
\begin{equation} \label{eq:ae_epme}
 \paw{g}_{mn\vb k,\nu\vb q} = \qty(\varepsilon_\nk - \varepsilon_{m\vb k + \vb q}) \braket{\pPsi_{m \vb k + \vb q} | \paw{S} | \partial_{\nu\vb q} \pPsi_\nk}
	\, ,
\end{equation}
as shown for instance in Ref.~\cite{nonrad}.
The difference between both approaches is rather subtle.
In the first case, one starts from the derivative of the full-potential orbital, then inserts the PAW transformation, \(\Psi_\nk = \pawT \ket{\pPsi_\nk} \), and finally applies the completeness relation to simplify the equations.
In our case, we first use the PAW transformation to phrase the problem in terms of a generalized eigenvalue problem~\cite{bloechl, kresse-PAW}.
The transformed problem does not have an explicit reference to the operator \(\pawT \).
We then calculate how the eigenvalues in the generalized eigenvalue problem change when ions are moved.
This approach does not involve explicit derivatives of the partial waves with respect to the ionic positions.

By expanding the AE orbitals according to Eq.~\eqref{eq:PAW_transform}, the following relation between both expressions can be found:
\begin{multline} \label{eq:ae_vs_paw_epme}
	g_{mn\vb k,\nu\vb q} - \paw{g}_{mn\vb k,\nu\vb q}
	\\ =
	\qty(\varepsilon_\nk - \varepsilon_{m\vb k + \vb q}) \braket{\pPsi_{m\vb k + \vb q} | \pawT^\dagger \qty(\partial_{\nu\vb q} \pawT) | \pPsi_\nk}
	\, .
\end{multline}
More details are presented in Appendix~\ref{sec:appendix:ae_vs_paw_epme}.
The difference between the two matrix elements is in general not zero, but notably, if \(\paw{S} = \id \), they coincide.

Furthermore, while the AE electron-phonon matrix is Hermitian, the same is not true for the one defined in Eq.~\eqref{eq:g_PAW}.
Hermitian conjugation, in this context, means the following:
\begin{equation}
	\qty(g_{mn\vb k, \nu \vb q})^\dagger
	\equiv
	g^*_{nm\vb k + \vb q, \nu -\vb q}
	\, .
\end{equation}
Since \(\partial^*_{\nu\vb q} = \partial_{\nu -\vb q} \), the Hermicity of the AE matrix element is easily established.
In the case of Eq.~\eqref{eq:g_PAW}, we obtain:
\begin{equation}
	\begin{split}
		\qty(\paw{g}_{mn\vb k, \nu \vb q})^\dagger
		& =
		\braket{\pPsi_{m\vb k + \vb q} | \partial_{\nu\vb q} \paw{H} - \varepsilon_{m\vb k + \vb q} \partial_{\nu\vb q} \paw{S} | \pPsi_\nk}
		\\ & \neq
		\paw{g}_{mn\vb k, \nu \vb q}
		\, .
	\end{split}
\end{equation}
This behavior is a direct consequence of dealing with a generalized eigenvalue problem.
Any physical observable, however, will involve terms of the form \(\norm{\paw{g}}^2 = \paw{g}\paw{g}^* \) which are self adjoint, hence, all observables are well defined and real.

As a matter of fact, it is also possible to calculate the FM and DW self energies using the AE matrix element~\eqref{eq:ae_epme}, e.g.~\cite{elphon_PAW_chaput}.
At first glance, this might seem like a reasonable proposal as it removes the necessity of dealing with non-Hermitian operators and perturbation theory for a generalized eigenvalue problem.
However, the matrix element \(\braket{\pPsi_{m\vb k + \vb q} | \pawT^\dagger \partial_{\nu\vb q} \pawT | \pPsi_\nk} \) that needs to be evaluated in this case contains explicit derivatives of PAW partial waves.
In our experience, this means that the results will be dependent on the completeness of the PAW partial waves, for instance, all summations should include core orbitals.
That being said, PAW potentials including many partial waves and treating many core states as valence orbitals (such as the GW PAW potentials) might enable accurate calculations of the AE electron-phonon matrix element using Eq.~\eqref{eq:ae_vs_paw_epme}.

The greatest merit in having defined the PAW matrix elements in Eq.~\eqref{eq:g_PAW} and~\eqref{eq:gS_PAW} is undoubtedly their fast convergence with respect to the included partial waves.
Furthermore, the matrix elements involving \(\paw{H} \), \(\paw{S} \) and their derivatives are already computationally available in most PAW implementations.
This is a significant advantage as it becomes possible to reuse existing routines.

\section{Wannier Interpolation}\label{eq:wannier_interpolation}
So far, a perturbative approach for calculating the phonon-induced electron self energy inside the PAW framework has been presented.
Additionally, it has been shown how an electron-phonon matrix element could be defined in this context.
In practical applications, it is often necessary to sample the first Brillouin zone densely, in turn requiring the calculation of a vast number of electron-phonon matrix elements.
When using a perturbative method such as DFPT directly, this approach can become very expensive.
In this section, an interpolation method based on Wannier functions that promises to reduce the computational cost is presented.

\subsection{Generalized Wannier Orbitals}

A generalized Wannier orbital, \(\ket{\mathcal{W}_{ml}} \), may be defined as
\begin{equation}
	\label{eq:wannier_from_bloch}
	\ket{\mathcal{W}_{al}} \equiv
	\frac{1}{\sqrt{N_\text{k}}} \sum_\nk
	\eul^{-\ci \vb k \vdot \vb* \xi_{la}}
	\ket{\Psi_\nk} U_{na,\vb k}
	\, ,
\end{equation}
such that its real-space wave function, \(\mathcal{W}_{al}\qty(\vb r) \equiv \braket{\vb r | \mathcal{W}_{al}} \), is localized~\cite{wannier-functions}.
In each cell, which is labeled by \(l \), multiple Wannier orbitals may exist that are distinguished by the Wannier index \(a \).
The vector \(\vb* \xi_{la} \) points towards the spatial center of the Wannier function \(\mathcal{W}_{al}\qty(\vb r) \).
\(N_\text{k} \) is the number of k-points in the first Brillouin zone.
The unitary matrix \(U_{na, \vb k} \), referred to as the Wannier transformation matrix, will be discussed below.
An inverse transformation can easily be specified, transforming a set of Wannier orbitals back into Bloch orbitals:
\begin{equation}
	\label{eq:bloch_from_wannier}
	\ket{\Psi_\nk} =
	\frac{1}{\sqrt{N_\text{k}}}
	\sum_{la}
	\eul^{\ci \vb k \vdot \vb* \xi_{la}}
	\ket{\mathcal{W}_{al}} U^\dagger_{an,\vb k}
	\, .
\end{equation}

Notably, the transformation matrices \(U_{na,\vb k} \) are non-unique.
A popular choice for them is the one that gives maximum localization with respect to a well-defined spread functional.
The resulting, so-called, maximally-localized Wannier functions need to be obtained from an iterative procedure~\cite{MLWF_iteration}.
Contrarily, in the present work, Wannier orbitals are utilized that are generated from a simple projection scheme such as the one used in reference~\cite{schueler-projection}, as detailed in the following section.

\subsection{Wannier Functions via Projection}

A simple way of constructing a set of symmetric Wannier functions suitable for interpolation is via projection.
To begin with, the Bloch manifold is projected onto a set of localized trial orbitals, \(\ket{\zeta_a} \):
\begin{equation}
	A_{na, \vb k} \equiv f_\nk \braket{\Psi_\nk | \zeta_a}
	\, .
\end{equation}
Let \(N_\text{B} \) be the number of Bloch orbitals involved in the projection and \(N_\mathcal{W} \) be the number of trial orbitals, then the dimension of the matrix \(A_{na, \vb k} \) is \(N_\text{B} \cross N_\mathcal{W} \).

The weight factors \(f_\nk \) provide a smooth cutoff for the Bloch orbitals as higher-lying states are assigned increasingly smaller weights.
This is important in the case of entangled bands, such as for metals, as opposed to an isolated manifold.
Including these weighting factors can substantially increase the suitability of the resulting Wannier orbitals for interpolation.
In practice, the \(f_\nk \) are modeled by a Fermi--Dirac distribution function, albeit with the ``Fermi'' energy at an appropriate point in the conduction band.

Generally, the rectangular projection matrix, \(A_{na, \vb k} \), has a rank not greater than the dimension of the trial orbitals' span.
Emphasis is put on the fact that the \(\ket{\zeta_a} \) need not be orthonormal.

The projection matrix is then expressed as a matrix product using a singular-value decomposition,
\begin{equation}
	A_{na, \vb k} = \sum_{m}^{N_\text{B}} \sum_{b}^{N_\mathcal{W}} X_{nm, \vb k} \Lambda_{mb, \vb k} Y^\dagger_{ba, \vb k}
	\, ,
\end{equation}
where \(X_{nm, \vb k} \) and \(Y_{ba, \vb k} \) are unitary square matrices with dimensions \(N_\text{B} \cross N_\text{B} \) and \(N_\mathcal{W} \cross N_\mathcal{W} \), respectively.
The matrix \(\Lambda_{mb, \vb k} \) has a rank \(\min\qty(N_\text{B}, N_\mathcal{W}) \), the same dimensions as \(A_{na, \vb k} \) and contains the latter's singular values on the main diagonal and zero everywhere else.

Finally, the Wannier transformation matrix is constructed using the unitary matrices obtained from the singular-value decomposition:
\begin{equation}
	U_{na, \vb k} =  \sum_{m}^{N_\text{B}} \sum_{b}^{N_\mathcal{W}} X_{nm, \vb k} \delta_{mb} Y^\dagger_{ba, \vb k}
	\, .
\end{equation}
One can easily verify that the resultant Wannier orbitals are orthonormal if the original Bloch orbitals are orthonormal.
Experience shows that Wannier orbitals constructed this way are suitable for Wannier interpolation~\cite{schueler-projection}.
This is discussed in the following section.

\subsection{Interpolation of the Electron-Phonon Matrix Element}\label{sec:Wannier_interpolation}

The interpolation scheme described in this section follows the same basic principles as outlined in references~\cite{giustino-coupling} and~\cite{mauri_interpol}.
Using Wannier interpolation, the electron-phonon matrix element can be obtained in the Bloch representation as
\begin{equation}
	\label{eq:g_interpol}
	\begin{split}
		g_{mn \vb k, \nu \vb q} & =
		\sum_{al} \sum_{bl'} \sum_{\kappa \alpha}
		\frac{1}{\sqrt{m_\kappa}}
		e_{\kappa \alpha, \nu \vb q}
		\\ & \cross
		\eul^{-\ci \qty(\vb k + \vb q) \vdot \qty(\vb* \xi_{la} - \vb R_{0 \kappa})}
		\eul^{\ci \vb k \vdot \qty(\vb* \xi_{l'b} - \vb R_{0 \kappa})}
		\\ & \cross
		U_{m a, \vb k + \vb q}
		g^\mathcal{W}_{albl',0 \kappa \alpha}
		U_{b n, \vb k}^\dagger
		\, ,
	\end{split}
\end{equation}
with
\begin{equation}
	\label{eq:g_Wannier}
	g^\mathcal{W}_{albl',p \kappa \alpha}
	\equiv
	\braket{\mathcal{W}_{al} | \partial_{p \kappa \alpha} \hat{H} | \mathcal{W}_{bl'}}
	\, .
\end{equation}
The equilibrium position of ion \(\kappa \) in cell \(l \) is denoted by \(\vb R_{l\kappa} \), and \(\vb* \xi_{la} \) is the center of the Wannier function.
Small differences to the original formalism in reference~\cite{giustino-coupling} exist deliberately to better match the present implementation.
For example, the translational invariance of the Wannier-space matrix element is used to confine the vibrational degrees of freedom to a single unit cell.
This is favorable in a finite difference scheme since the number of required independent displacements is minimized.

The cell indices \(l \) and \(l' \) corresponding to the involved Wannier functions go over the entire lattice.
If the Wannier-space matrix element, \(g^\mathcal{W} \), decays fast enough as a function of \(\abs{\vb R_{0\kappa} - \vb* \xi_{la}} \), however, it is possible to introduce an effective cutoff radius.
Matrix elements beyond this cutoff are assumed to be zero and the interpolation can be performed efficiently using small matrices.

At arbitrary k-points, the interpolated electronic eigenvalues can be obtained by diagonalization of the Wannier-interpolated Hamiltonian matrix,
\begin{equation}
	\label{eq:H_interpol_1}
	H^\mathcal{W}_{ab\vb k}
	\equiv
	\sum_{l} \eul^{-\ci \vb k \vdot \qty(\vb* \xi_{la} - \vb* \xi_{0b})}
	\braket{\mathcal{W}_{al} | \hat{H} | \mathcal{W}_{b0}}
	\, ,
\end{equation}
\emph{i.e.}, by determining unitary matrices, \(U_{ma, \vb k} \), that diagonalize \(H^\mathcal{W}_{ab\vb k} \):
\begin{equation}
	\label{eq:H_interpol_2}
	\delta_{mn} \varepsilon_\nk =
	\sum_{ab}
	U_{ma, \vb k}
	H^\mathcal{W}_{ab\vb k}
	U^\dagger_{bn, \vb k}
	\, .
\end{equation}
Equations~\eqref{eq:H_interpol_1} and~\eqref{eq:H_interpol_2} together also allow to interpolate the electronic band structure to arbitrary Bloch vectors.

\subsection{Wannier Interpolation in the PAW Framework}

In the PAW method, Eq.~\eqref{eq:wannier_from_bloch} is expressed in terms of pseudo orbitals,
\begin{equation}
	\ket{\paw{\mathcal{W}}_{al}} \equiv
	\frac{1}{\sqrt{\Omega_\text{BZ}}} \bzint \dd[3]{k}
	\eul^{-\ci \vb k \vdot \vb* \xi_{la}}
	\ket{\pPsi_\nk} U_{na,\vb k}
	\, ,
\end{equation}
which defines a set of pseudo Wannier orbitals, \(\ket{\paw{\mathcal{W}}_{al}} \), completely analogous to the AE case.
In other words, the Wannier transformation does not affect the atom-specific part of the PAW transformation.

A generalization of Wannier interpolation to the PAW method is then, in principle, straightforward.
For example, to interpolate the electronic band structure in the PAW method, one simply replaces the AE orbitals and operators in Eq.~\eqref{eq:H_interpol_1} with their pseudo counterparts:
\begin{equation}
	\braket{\mathcal{W}_{al} | \hat{H} | \mathcal{W}_{b0}} =
	\braket{\paw{\mathcal{W}}_{al} | \paw{H} | \paw{\mathcal{W}}_{b0}}
	\, .
\end{equation}

In the case of the electron-phonon matrix element, the matrix elements involving \(\partial_{\nu\vb q} \paw{H} \) and \(\partial_{\nu\vb q} \paw{S} \) can be interpolated separately.
The required band energies, \(\varepsilon_\nk \), can be obtained from Eq.~\eqref{eq:H_interpol_1} and~\eqref{eq:H_interpol_2}.

\section{Results and Discussion} \label{sec:results}

Before the Wannier-Interpolation scheme discussed in Sec.~\ref{sec:Wannier_interpolation} can be used in practice, a few conditions must be met.
In particular, the Wannier-space matrix elements involving the PAW Hamiltonian, its derivative and the derivative of the PAW overlap operator need to spatially decay sufficiently fast so that the sums in Eq.~\eqref{eq:g_interpol} and~\eqref{eq:H_interpol_1} converge.
Additionally, the set of Wannier functions used for interpolation must closely match the electronic character of the target range of bands included in the transformation.
It is important to verify these criteria independently for each material.
Otherwise, the quality of the interpolation might deteriorate.

In this section, attention is initially placed on diamond, owing mostly to its extensive coverage in literature.
Afterwards, phonon-induced band-gap renormalizations are calculated for a set of semiconductors.
We also take the opportunity to compare our results with the ones obtained by Karsai~\cite{one-shot-Karsai} using a stochastic one-shot method.
The latter has recently also been implemented in VASP providing a solid ground for comparison.

\subsection{Diamond}

In diamond, the Wannier-projection process is straight forward.
The four highest valence bands and the four lowest conduction bands can be spanned by one s-like and three p-like Wannier orbitals on each atom.
With two atoms per primitive cell, the number of bands that can be spanned is eight.
Four of which are in the valence band, while the other four are in the conduction band.

All calculations on diamond are performed using the PBE~\cite{GGA-PBE-1,GGA-PBE-2} PAW potential with the default electronic cutoff of \(\SI{400}{\electronvolt} \).
The lattice parameter is set to the experimental value of \(\SI{3.567}{\angstrom} \)~\cite{exp_lattice_C}.

\begin{figure*}[t]
	\includegraphics{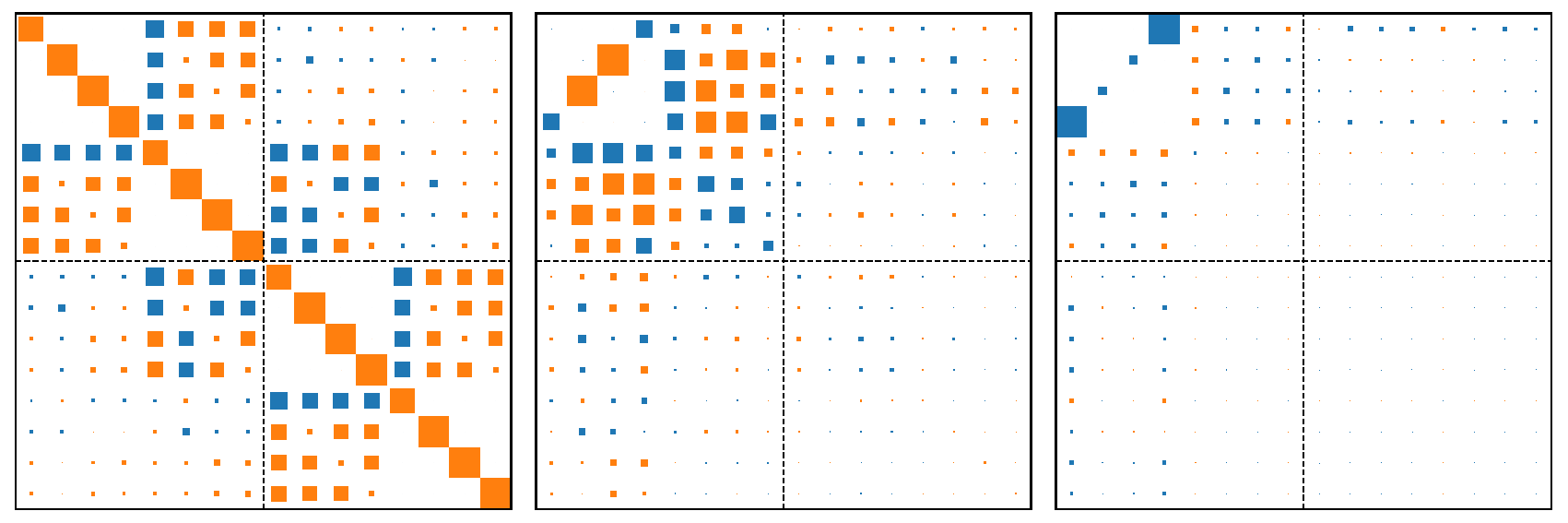}
	\caption{
		Numerical structure of various Wannier matrices for diamond.
		The three figures show parts of the real-valued matrices \(\braket{\paw{\mathcal{W}}_{al} | \paw{H} | \paw{\mathcal{W}}_{bl'}} \) (left),
		\(\braket{\paw{\mathcal{W}}_{al} | \partial_{000} \paw{H} | \paw{\mathcal{W}}_{bl'}} \) (middle) and \(\braket{\paw{\mathcal{W}}_{al} | \partial_{000} \paw{S} | \paw{\mathcal{W}}_{bl'}} \) (right).
		The dashed lines separate regions corresponding to different unit cells (indices \(l \) and \(l' \)), while the rows and columns of colored squares correspond to the different Wannier orbitals (indices \(a \) and \(b \); in this case, 1 s-like and 3 p-like orbitals per atom).
		The area of each square is proportional to the absolute value of the corresponding matrix element, while the color indicates positive (orange) and negative (blue) values.
	}
	\label{fig:wannier_mats_structure}
\end{figure*}

Small sections of the Wannier-space matrices of the Hamiltonian, its derivative and the derivative of the PAW overlap are visualized in Fig.~\ref{fig:wannier_mats_structure} (the unperturbed overlap matrix is the identity matrix).
The matrix elements corresponding to the atom at which the perturbation occurs are located in the first four rows (or columns) starting from the top left corner of the shown section.
One can clearly see that, after taking the derivative with respect to that atom, only matrix elements involving atoms around that perturbation have significant contributions.
This supports the claim that the real-space sums appearing in Eq.~\eqref{eq:g_interpol} can be truncated at a fairly small radius without jeopardizing their convergence.

\begin{figure}[h]
	\includegraphics{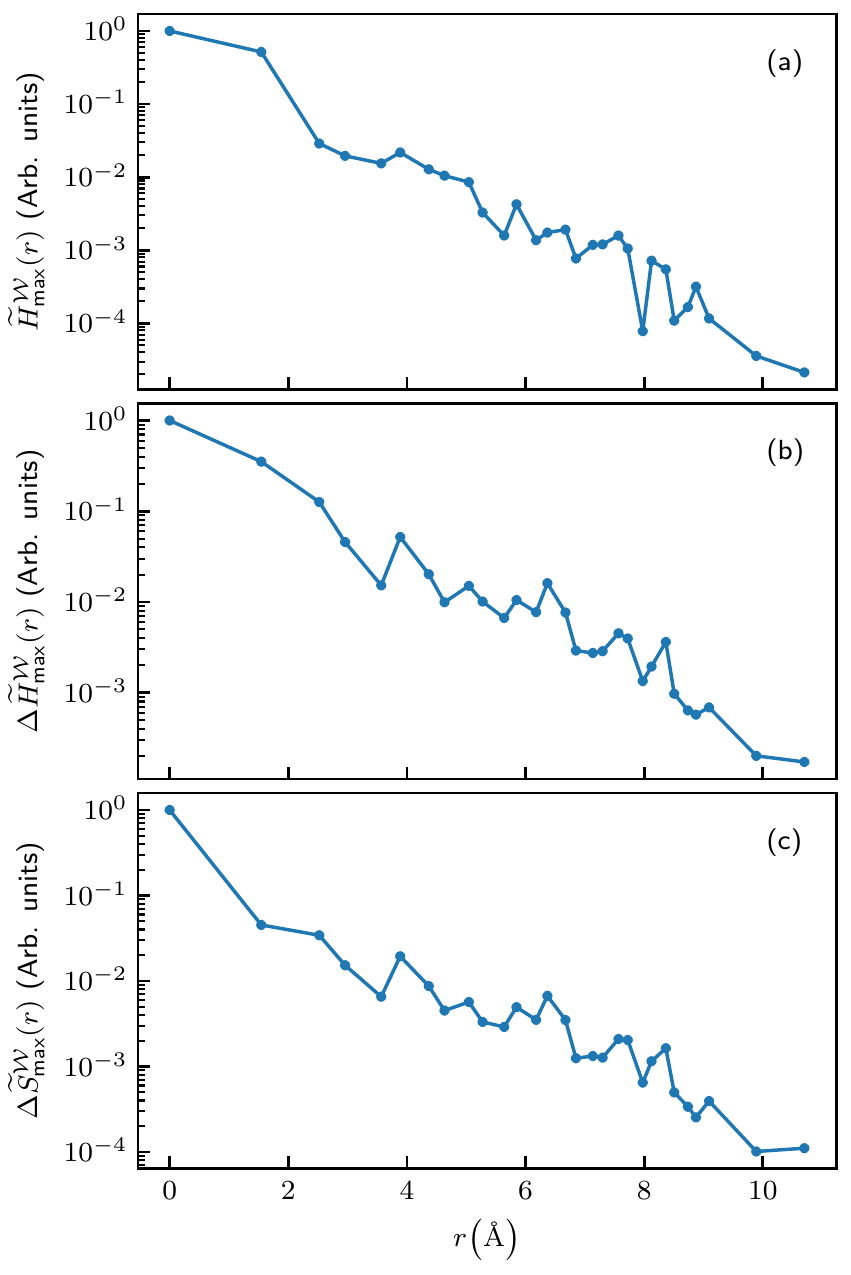}
	\caption{
		Various Wannier-space matrix elements for diamond plotted as functions of the distance between the corresponding Wannier centers.
		For each possible distance, only the largest absolute value is shown.
		A logarithmic scale is used to highlight the exponential spatial decay.
		Panel (a) shows the unperturbed PAW Hamiltonian matrix, \(\paw{H}^\mathcal{W}_\text{max} \qty(r) \equiv \max_\qty{al} \braket{\paw{\mathcal{W}}_{al} | \paw{H} | \paw{\mathcal{W}}_{00}} \), with \(\abs{\vb* \xi_{la} - \vb* \xi_{00}} = r \).
		Panel (b) shows the Hamiltonian matrix for a particular mono-atomic perturbation, \(\Delta \paw{H}^\mathcal{W}_\text{max} \qty(r) \equiv \max_\qty{al} \braket{\paw{\mathcal{W}}_{al} | \partial_{000} \paw{H} | \paw{\mathcal{W}}_{00}} \), with \(\abs{\vb* \xi_{la} - \vb R_{00}} = r \).
		Finally, (c) is the same as (b) but showing the perturbed PAW overlap matrix.
	}
	\label{fig:C_decay}
\end{figure}

A meaningful way of quantifying the spatial decay of these matrix elements is to study their absolute values as a function of distance between the associated Wannier centers in the periodic supercell.
Figure~\ref{fig:C_decay} shows this dependency on distance for the Hamiltonian, its derivative and the derivative of the PAW overlap operator.
Clearly, the data decay roughly exponentially with the distance.

Finally, we would like to showcase the importance of the supercell size and the number of q-points in the convergence of an observable.
The size of the supercell is relevant because the number of Wannier orbitals is directly tied to the number of atoms in our implementation.
Therefore, the Wannier-space cutoff radius is implicitly given by the size of the supercell and is implemented using a minimum-image convention within the periodic supercell.
In our studies, we use \(4\cross 4\cross 4 \), \(5\cross 5\cross 5 \) and \(6\cross 6\cross 6 \) supercells comprised of primitive unit cells, each containing two atoms.

We calculate the zero-point renormalization (ZPR) of the direct band gap of diamond using Eq.~\eqref{eq:FM_self_energy} and~\eqref{eq:DW_self_energy}.
The Brillouin-zone integrals are approximated by sums over a dense q-point mesh.
Here, we simply choose the mesh to contain the q-points commensurate with the supercell and subdivide it in all three reciprocal-lattice directions in order to increase the q-point density.

\begin{figure}[h]
	\includegraphics{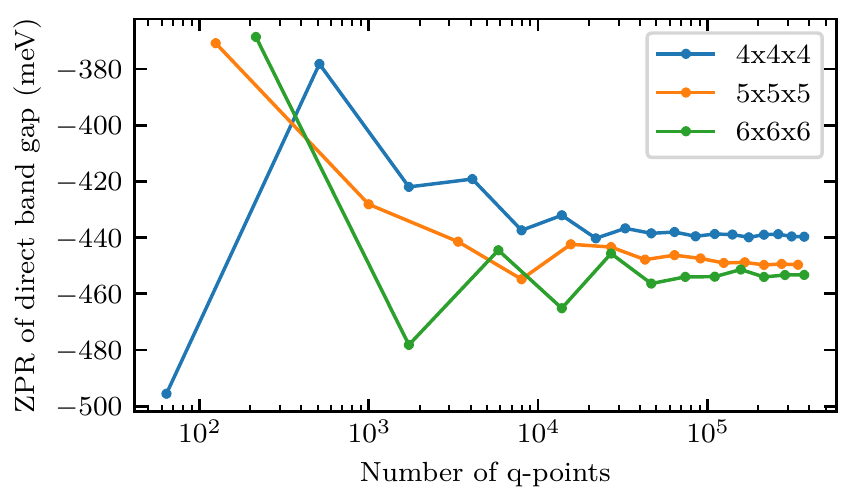}
	\caption{
		Convergence of the renormalization of the direct band gap of diamond as a function of cell size and q-point density.
		The number of q-points is determined as the number of q-points commensurate with the super cell times the number of subdivisions in one direction cubed.
	}
	\label{fig:ZPR_convergence}
\end{figure}

The ZPR of the direct gap as a function of the Brillouin-zone sampling density is shown in Fig.~\ref{fig:ZPR_convergence} for different supercell sizes.
Based on these calculations, the most well-converged result for the ZPR of the direct gap of diamond is \(\SI{-427}{\milli\electronvolt} \) for the \(6\cross 6\cross 6 \) supercell.

Judging from the available data, we conclude that cell sizes beyond a \(6\cross 6\cross 6 \) supercell would not significantly change the result for diamond.
We also note that q-point convergence is reached independent of the original cell size at around \(10^5 \)~q-points inside the first Brillouin zone.

\begin{figure}[h]
	\includegraphics{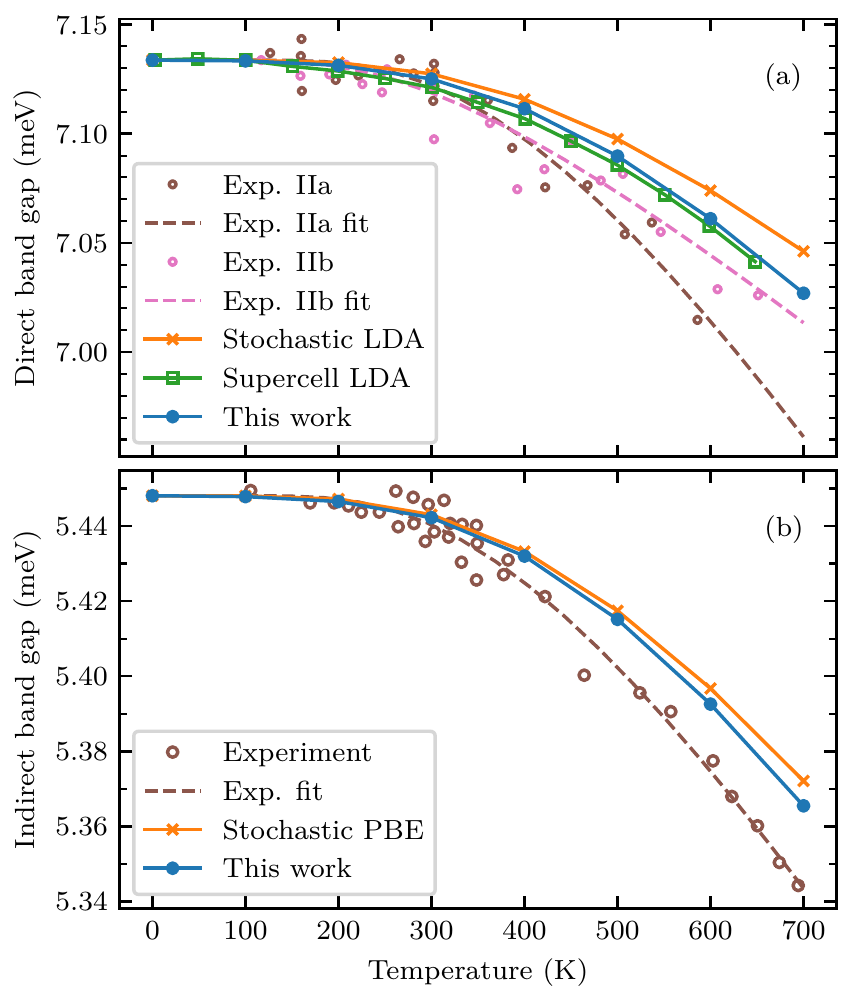}
	\caption{
		Temperature dependence of the direct and indirect band gap of diamond, with the present results obtained from a \(5\cross 5\cross 5 \) supercell.
		The data corresponding to the stochastic one-shot method (orange crosses) were taken from~\cite{one-shot-Karsai}, while the ones for a reference super-cell calculation (green squares) were extracted from~\cite{gonze-temperature-bandstruct}.
		The experimental data for the direct gap correspond to samples IIa and IIb in~\cite{exp-temperature-bandstruct-direct} and for the indirect gap, the data are taken from~\cite{exp-temperature-bandstruct-indirect}.
		Experimental data sets are fitted with the analytic function proposed in~\cite{exp-temperature-bandstruct-indirect}.
		All curves are shifted as to coincide with the fitted experimental data at zero temperature (in panel (a), with respect to data set IIa).
	}
	\label{fig:temperature}
\end{figure}

Finally, we also calculate the temperature-dependent direct and indirect band gap of diamond using a \(5\cross 5\cross 5 \) supercell at fixed volume.
The results are shown in Fig.~\ref{fig:temperature}.
For the sake of comparison, experimental as well as other ab-initio results are included.
The experimental data sets are fitted with an analytical function advocated in~\cite{exp-temperature-bandstruct-indirect}, claimed to be superior to fits based on the Varshni equation~\cite{varshni-equation}.
The ab-initio data from Karsai correspond to one-shot calculations for a \(5\cross 5\cross 5 \) supercell using density-functional theory (extracted from the LDA and PBE data sets in Fig.~3 and~5 in~\cite{one-shot-Karsai}, respectively).
Lastly, the data set produced by Gonze \etal~\cite{gonze-temperature-bandstruct}, using the same underlying methodology as presented here, is shown.

The results are satisfactory as there is good agreement with the other ab-initio data sets.
In the case of the direct gap, our results tend to agree quite well with the results published by Gonze \etal.
This is to be expected since the underlying computational methods are essentially the same.
The deviation from the experimental data at high temperatures is a result of the neglect of the volume expansion as a function of the temperature (quasi-harmonic effects) as discussed in reference~\cite{one-shot-Karsai}.

\subsection{Band-gap Renormalization in other Materials}

The present algorithm has also been employed to calculate the ZPR of band gaps for AlAs, AlP, AlSb, BN, C, GaN, GaP, Si and SiC.
These materials share the same zincblende (or diamond) structure but with different lattice parameters.
Appropriate Wannier orbitals can again be obtained by a projection on one s-like and three p-like orbitals.
Again, PBE potentials and default electronic cutoff radii are used and ZPRs corresponding to degenerate bands are averaged.
The computational parameters are summarized in table~\ref{tab:parameters}.
\begin{table}
	\begin{ruledtabular}
		\begin{tabular}{lccc}
			&
			\begin{tabular}[c]{@{}c@{}} Lattice \\ parameter \\ (\si{\angstrom}) \end{tabular} &
			\begin{tabular}[c]{@{}c@{}} Plane-wave \\ cutoff \\ (\si{\electronvolt}) \end{tabular} &
			Reference
			\\
		\colrule
			AlAs	& 5.661	& 240 & \cite{exp_lattice_AlAs}	\\
			AlP		& 5.463	& 255 & \cite{exp_lattice_AlP}	\\
			AlSb	& 6.136	& 240 & \cite{exp_lattice_AlSb}	\\
			BN		& 3.616	& 400 & \cite{exp_lattice_BN}	\\
			C		& 3.567	& 400 & \cite{exp_lattice_C}	\\
			GaN		& 4.535	& 400 & \cite{one-shot-Karsai}	\\
			GaP		& 5.451	& 255 & \cite{exp_lattice_GaP}	\\
			Si		& 5.431	& 245 & \cite{exp_lattice_Si}	\\
			SiC		& 4.358	& 400 & \cite{exp_lattice_SiC}	\\
		\end{tabular}
	\end{ruledtabular}
	\caption{
		List of computational parameters for the simulated materials.
	}
	\label{tab:parameters}
\end{table}
The lattice parameters are chosen in accordance with the ones in~\cite{one-shot-Karsai} in order to maximize the comparability of the data.

With the exception of GaN, all of these materials feature an indirect band gap.
In order to determine the precise location of the band transition inside the first Brillouin zone, Wannier-interpolation is used to calculate the electronic band structure along finely sampled high-symmetry lines.
These data are then scanned for the conduction-band minimum and the valence-band maximum.
The total ZPR is then the difference between the respective energy shifts.

Calculations are performed for \(4\cross 4 \cross 4 \) and \(5\cross 5\cross 5 \) supercells.
All results are converged with respect to the number of q-points.
Table~\ref{tab:ZPR_results} shows the final results for the aforementioned materials.
\begin{table}[]
	\begin{ruledtabular}
		\begin{tabular}{lcccc}
			&
			\begin{tabular}[c]{@{}c@{}} ZPR \\ (\(4\cross 4\cross 4 \)) \end{tabular} &
			\begin{tabular}[c]{@{}c@{}} ZPR \\ (\(5\cross 5\cross 5 \)) \end{tabular} &
			Theory &
			Exp.
			\\
		\colrule
			AlAs	& -0.048	& -0.056	& --		& --		\\
			AlP		& -0.057	& -0.065	& --		& --		\\
			AlSb	& -0.032	& -0.039	& --		& -0.039	\\
			BN		& -0.294	& -0.321	& -0.262	& --		\\
					&			&			& -0.331	&			\\
			C		& -0.344	& -0.368	& -0.330	& -0.340	\\
					&			&			& -0.343	& -0.370	\\
					&			&			& -0.379	&			\\
			GaN		& -0.090	& -0.095	& -0.127	& --		\\
			GaP		& -0.049	& -0.049	& --		& --		\\
			Si		& -0.044	& -0.055	& -0.058	& -0.050	\\
					&			&			& -0.060	& -0.064	\\
					&			&			& -0.064	&			\\
			SiC		& -0.113	& -0.120	& -0.109	& --		\\
		\end{tabular}
	\end{ruledtabular}
	\caption{
		Band-gap renormalization due to zero-point lattice vibrations for a selection of materials obtained via the presented algorithm.
		All results are converged with respect to the number of q-points.
		Supercell sizes are given in parentheses.
		Theoretical literature values are presented for BN~\cite{result_theory_BN}, C~\cite{result_theory_C,result_theory_C_Si_A_NA}, GaN~\cite{result_theory_GaN}, Si~\cite{result_theory_Si,result_theory_C_Si_A_NA} and SiC~\cite{result_theory_SiC}.
		In addition, experimental results are presented for AlSb~\cite{result_exp_Si_AlSb}, C~\cite{result_exp_Si_C, result_exp_C} and Si~\cite{result_exp_Si_AlSb,result_exp_Si_C}.
		Energies are in \si{\electronvolt}.
	}
	\label{tab:ZPR_results}
\end{table}

A closer look at the data reveals that the ZPR increases with cell size, lowering the band gap, except for GaP where the ZPR stays the same.
This is consistent with the convergence behavior shown in Fig.~\ref{fig:ZPR_convergence}, indicating that the \(4\cross 4\cross 4 \) supercell is insufficiently large to reach convergence.
The mean absolute difference in the ZPRs between the \(4\cross 4 \cross 4 \) and \(5\cross 5\cross 5 \) cell is about \SI{11}{\milli\electronvolt}, which also agrees favorably with the results obtained for the direct gap of diamond.

Finally, we compare our ZPR results with the ones from Ref.~\cite{one-shot-Karsai}, which is summarized in table~\ref{tab:ZPR_compare}.
\begin{table}[]
	\begin{ruledtabular}
		\begin{tabular}{lccccc}
			&
			\multicolumn{2}{c}{\(4\cross 4\cross 4 \)} &
			\multicolumn{2}{c}{\(5\cross 5\cross 5 \)} &
			\(5\cross 5\cross 5 \)
			\\
			& AHC & One shot & AHC & One shot & Converged \\
		\colrule
			AlAs	& -0.050	& -0.051	& -0.055	& -0.063	& -0.056	\\
			AlP		& -0.060	& -0.062	& -0.067	& -0.070	& -0.065	\\
			AlSb	& -0.041	& -0.050	& -0.039	& -0.043	& -0.039	\\
			BN		& -0.257	& -0.269	& -0.290	& -0.294	& -0.321	\\
			C		& -0.396	& -0.363	& -0.337	& -0.320	& -0.368	\\
			GaN		& -0.086	& -0.102	& -0.095	& -0.094	& -0.095	\\
			GaP		& -0.047	& -0.072	& -0.044	& -0.057	& -0.049	\\
			Si		& -0.054	& -0.064	& -0.054	& -0.065	& -0.055	\\
			SiC		& -0.104	& -0.104	& -0.130	& -0.120	& -0.120	\\
		\end{tabular}
	\end{ruledtabular}
	\caption{
		Zero-point band-gap renormalizations calculated via the presented AHC algorithm compared to ones calculated with the one-shot method.
		Results for the one-shot method are taken from table 3 in~\cite{one-shot-Karsai}, with the exception of SiC.
		Only q-points commensurate with the respective supercell are included to maximize the comparability of the two methods.
		The converged results from table~\ref{tab:ZPR_results} are shown once more in the last column for the sake of comparison.
		Energies are in \si{\electronvolt}.
	}
	\label{tab:ZPR_compare}
\end{table}
Since the data sets are generated by fundamentally different methods, it is desirable to tune the computational parameters in order to maximize comparability.
The lattice parameters are already chosen to be identical and the PBE potential is used in both cases.
In the one-shot method, only phonon modes at the \(\Gamma \)-point of the supercell are included in the calculation.
This is equivalent to sampling only q-points in the primitive cell's Brillouin zone that are commensurate with the supercell.
Therefore, the integrals appearing in Eq.~\eqref{eq:FM_self_energy} and~\eqref{eq:DW_self_energy} are restricted to sums over these commensurate q-points in the second and fourth column of table~\ref{tab:ZPR_compare}.
This restriction also applies to finding the location of the conduction-band minimum.

Even with these modifications, there are still fundamental differences in the methodology that cannot be easily accounted for.
The one-shot method implicitly includes anharmonic contributions that are completely absent from second-order perturbation theory.
Moreover, AHC theory relies on the rigid-ion approximation to evaluate the DW self energy, which induces errors.
Probably most problematic, however, is the issue of convergence with respect to the number of virtual orbitals that arises in methods based on Wannier interpolation.
In principle, one needs to include all occupied and unoccupied orbitals in sums over states, such as the ones appearing in Eq.~\eqref{eq:FM_self_energy} and~\eqref{eq:DW_self_energy}.
The set of Wannier functions only spans a small subspace of the full Hilbert space.
This has potential ramifications for the accuracy of the method.
Unfortunately, it is very difficult to estimate bounds on the error induced by neglecting higher-lying states in the perturbational calculations.
A numerical comparison with a different method integrated in the same software package is obviously valuable.

We need to report that the values for the ZPR of SiC as listed in reference~\cite{one-shot-Karsai} are erroneous, as subsequent calculations have shown.
The values presented here have been recalculated using the one-shot method yielding \SI{-104}{\milli\electronvolt} and \SI{-120}{\milli\electronvolt} for \(4\cross 4\cross 4 \) and \(5\cross 5\cross 5 \) supercells, respectively.

The agreement between both data sets is remarkably good, especially in light of the aforementioned differences in methodology.
The mean relative error between the data sets is about \SI{11}{\percent} for the smaller cell while for the larger cell it is approximately \SI{9}{\percent}.

Generally, the one-shot method yields systematically larger absolute band-gap renormalizations (with the exception of C and SiC).
We relate this to the fact that the one-shot method implicitly accounts for excitations in all unoccupied states, whereas in the present perturbational calculations, we limit the calculations to 8 bands per unit cell.

We relate this to the fact that the one shot-method does not require a summation over states, and in the present perturbational calculations, we truncate the calculation to 8 bands per unit cell.
We note that perturbation theory is not variational, implying that the inclusion of more bands does not necessarily increase the ZPR, although usually this is the case.

\section{Conclusion} \label{sec:conclusion}

In this paper, the zero-point and finite-temperature renormalizations of the electronic bands due to electron-phonon interactions are derived and determined in the PAW framework.
As usual in the AHC method, we define the electron-phonon matrix element as the second derivative of the one-electron energies in the adiabatic (Born--Oppenheimer) approximation, \emph{i.e.}, neglecting explicit frequency dependencies of the external perturbation. 
Since the PAW method yields a generalized eigenvalue problem, the second derivatives of the eigenvalues also involve an eigenvalue-dependent ``non-Hermitian'' matrix element.
This matrix element lacks some symmetries that are usually present in all-electron formulations, and we have discussed this issue at some length.

An important characteristic of the present method is that the actual implementation does not rely on linear-response theory.
Instead, the first-order change of the orbitals is determined by small finite displacements.
This makes the method broadly applicable to any functional, including hybrid functionals as well as more complicated exchange-correlation functionals, where higher-order derivatives of the functionals are not readily computable.
Finite differences, however, also imply the use of supercells.
To mitigate this problem and to reliably determine the linear response of the orbitals, the orbitals are transformed to a Wannier representation using a projection scheme.
This has the added advantage that a Wannier interpolation of the electron-phonon matrix elements is readily possible.
This interpolation is, as a matter of fact, again adapted to the PAW method.

Using the developed method, numerical calculations are performed on diamond and a selection of binary compounds.
To begin with, convergence tests are performed on diamond showing that a \(5\cross 5\cross 5 \) supercell and the \(\Gamma \)-point suffice to determine reliable electron-phonon matrix elements.
These can then be interpolated to much larger supercells and thus much denser wave-vector grids.
We find that interpolation to about a million q-points suffices for results converged to few meV for the ZPR.
The temperature-dependence of the direct and indirect band gaps of diamond are then studied.
Our findings follow the same trends as other computational studies, underestimating the temperature dependence of the band-gap renormalization (this has previously been related to the neglect of the thermal expansion~\cite{one-shot-Karsai} and this previous assessment remains unchallenged).
In addition, band-gap renormalizations have been calculated for a set of binary compounds.
The results obtained by Wannier interpolation to many q-points compare very well with the available computational and experimental results found in the literature.
A comparison of the one-shot method with the values obtained without interpolation to a finer q-point grid shows excellent agreement for the band-gap renormalization at wave vectors commensurate with the super cell.
Not only does the excellent numerical agreement between the different methods serve as a proof-of-concept for the two implementations, it also shows that, within the set of tested materials, the employed approximations of all methods seem to be well justified.
Typically, errors introduced by using a finite supercell, by neglecting interpolation to a fine wave-vector grid, or by the rigid-ion approximation all amount to about 10\% in total.
Hence, if absolute accuracy is not an issue, most methods will yield reasonably reliable results.
Specifically, our present implementation of the Wannier-interpolation method gives good estimates of the band-structure renormalization at zero and finite temperature compared to other codes, while taking full advantage of the numerical convenience of the PAW method.

As an outlook, it is important to remember that one of the strengths of the Wannier-interpolation method lies in the efficient calculation of the imaginary part of the electron self energy.
Future endeavors will focus on a PAW formulation of the fully frequency-dependent self energy in second quantization.
In addition, the inclusion of phonon-induced long-range dielectric effects (Fr\"ohlich-like terms) that prominently occur in polar materials will be necessary~\cite{verdi-polar,sjakste-polar} in order to obtain accurate results for polar materials.

\section{Acknowledgements}

We gratefully acknowledge financial support provided by the Austrian Science Fund (FWF) No. I2460-N36.

\appendix

\section{Derivation of the First and Second Derivatives of the Kohn--Sham Eigenvalues} \label{sec:appendix:FM_DW_contributions}

This section details the steps leading up to Eq.~\eqref{eq:FM_contribution} and~\eqref{eq:DW_contribution}.
First though, it is important to note that the pseudo orbitals, \(\ket{\pPsi_\nk} \), are \(\paw{S} \)-orthonormal,
\begin{equation}
	\label{eq:appendix:orthogonality_constraint}
	\braket{\pPsi_{m\vb k} | \paw{S} | \pPsi_{n \vb k'}} = \delta_{mn} \delta_{\vb k \vb k'}
	\, ,
\end{equation}
and fulfill the following completeness relations:
\begin{align}
	\id & =
	\sum_\nk \pawT \dyad*{\pPsi_\nk}{\pPsi_\nk} \pawT^\dagger
	\, ,	\\ \id & =
	\sum_\nk \paw{S} \dyad*{\pPsi_\nk}{\pPsi_\nk}
	\, ,\\ \id & =
	\label{eq:appendix:completeness_3}
	\sum_\nk \dyad*{\pPsi_\nk}{\pPsi_\nk} \paw{S}
	\, .
\end{align}

To begin with, the expressions for the first and second derivatives of the KS eigenvalues are derived starting from Eq.~\eqref{eq:PAW_KS_equation}.
Applying the derivative once yields
\begin{equation}
	\partial_\tau \qty(\paw{H} \ket{\pPsi_\nk} - \varepsilon_\nk \paw{S} \ket{\pPsi_\nk}) = 0
	\, ,
\end{equation}
which is evaluated using the chain rule:
\begin{equation}
	\label{eq:appendix:KS_first_derivative_beginning}
	\begin{split}
		\partial_\tau \varepsilon_\nk \paw{S} \ket{\pPsi_\nk}
		& =
		\qty(\partial_\tau \paw{H} - \varepsilon_\nk \partial_\tau \paw{S}) \ket{\pPsi_\nk} \\
		& +
		\qty(\paw{H} - \varepsilon_\nk \paw{S}) \ket{\partial_\tau \pPsi_\nk}
		\, .
	\end{split}
\end{equation}
Multiplication from the left with the state vector \(\bra{\pPsi_\nk} \) results in
\begin{equation}
	\begin{split}
		\partial_\tau \varepsilon_\nk \overbrace{\braket{\pPsi_\nk | \paw{S} | \pPsi_\nk}}^{= 1}
		& =
		\braket{\pPsi_\nk | \partial_\tau \paw{H} - \varepsilon_\nk \partial_\tau \paw{S} | \pPsi_\nk} \\
		& +
		\underbrace{\bra{\pPsi_\nk} \paw{H} - \varepsilon_\nk \paw{S}}_{= 0} \ket{\partial_\tau \pPsi_\nk}
		\, ,
	\end{split}
\end{equation}
where the Hermitian conjugate of the generalized KS equation has been used to cancel the last term.
Finally, the first derivative of the KS eigenvalue reads:
\begin{equation}
	\label{eq:appendix:KS_first_derivative}
	\partial_\tau \varepsilon_\nk
	=
	\braket{\pPsi_\nk | \partial_\tau \paw{H} - \varepsilon_\nk \partial_\tau \paw{S} | \pPsi_\nk}
	\, .
\end{equation}
Following suit, the second derivative is evaluated by differentiating Eq.~\eqref{eq:appendix:KS_first_derivative}:
\begin{equation}
	\partial_\tau \partial_{\tau'} \varepsilon_\nk =
	\partial_\tau \qty(\braket{\pPsi_\nk | \partial_{\tau'} \paw{H} - \varepsilon_\nk \partial_{\tau'} \paw{S} | \pPsi_\nk})
	\, .
\end{equation}
Once again, the chain rule is applied:
\begin{equation}
	\label{eq:appendix:KS_second_derivative_beginning}
	\begin{split}
		\partial_\tau \partial_{\tau'} \varepsilon_\nk
		& =
		\braket{\pPsi_\nk | \partial_\tau \partial_{\tau'} \paw{H} - \varepsilon_\nk \partial_\tau \partial_{\tau'} \paw{S} | \pPsi_\nk}
		\\ & +
		\braket{\partial_\tau \pPsi_\nk | \partial_{\tau'} \paw{H} - \varepsilon_\nk \partial_{\tau'} \paw{S} | \pPsi_\nk}
		\\ & +
		\braket{\pPsi_\nk | \partial_{\tau'} \paw{H} - \varepsilon_\nk \partial_{\tau'} \paw{S} | \partial_\tau \pPsi_\nk}
		\\ & -
		\partial_\tau \varepsilon_\nk \braket{\pPsi_\nk | \partial_{\tau'} \paw{S} | \pPsi_\nk}
		\, .
	\end{split}
\end{equation}
The term \(\partial_\tau \varepsilon_\nk \) is already known and can be substituted back in from Eq.~\eqref{eq:appendix:KS_first_derivative}.
In order to avoid the derivative acting directly on the state vectors, the PAW completeness relation can be used.
Subsequently, a suitable expression for \(\ket{\partial_\tau \pPsi_\nk}  \) is derived from which the final result will follow.

One may begin with Eq.~\eqref{eq:appendix:KS_first_derivative_beginning}.
Under the constraint \(\qty(m \vb k') \neq \qty(\nk) \), multiplication with \(\bra{\pPsi_{m \vb k'}} \) from the left yields:
\begin{multline}
	\partial_\tau \varepsilon_\nk \overbrace{\braket{\pPsi_{m\vb k'} | \paw{S} | \pPsi_\nk}}^{= 0}
	=
	\braket{\pPsi_{m\vb k'} | \partial_\tau \paw{H} - \varepsilon_\nk \partial_\tau \paw{S} | \pPsi_\nk}
	\\ +
	\underbrace{\bra{\pPsi_{m\vb k'}} \paw{H} - \varepsilon_\nk \paw{S}}_{\mathclap{= \qty(\varepsilon_{m\vb k'} - \varepsilon_\nk) \bra{\pPsi_{m\vb k'}} \paw{S}}} \ket{\partial_\tau \pPsi_\nk}
	\, .
\end{multline}
At this point, it is assumed that all electronic bands are non degenerate (\(\varepsilon_\nk - \varepsilon_{m \vb k'} \neq 0 \)), resulting in:
\begin{equation}
	\label{eq:appendix:state_derivative_intermediate}
	\braket{\pPsi_{m\vb k'} | \paw{S} | \partial_\tau \pPsi_\nk}
	=
	\frac{\braket{\pPsi_{m \vb k'} | \partial_\tau \paw{H} - \varepsilon_\nk \partial_\tau \paw{S} | \pPsi_\nk}}{\varepsilon_\nk - \varepsilon_{m \vb k'}}
	\, .
\end{equation}
The final step in isolating \(\ket{\partial_\tau \pPsi_\nk} \) now involves using the completeness relation in Eq.~\eqref{eq:appendix:completeness_3} and summing over all states \(\ket{\Psi_{m \vb k'}} \).
In the AE case, this conveniently gives the final result since the term with \(\qty(m \vb k') = \qty(\nk) \) would not contribute to the sum, owing to the fact that \(\braket{\Psi_\nk | \partial_\tau \Psi_\nk} = 0 \).
In the PAW method, the term \(\braket{\pPsi_\nk | \paw{S} | \partial_\tau \pPsi_\nk} \) is, however, in general non zero.
Therefore, completing the sum must explicitly account for that contribution.
In order to compactly write the final result, the shorthand notations introduced in Eq.~\eqref{eq:shorthand_h} and~\eqref{eq:shorthand_hS} as well as the primed-sum notation for omitting the \(\qty(m \vb k') = \qty(\nk) \) case are used thereafter.

Multiplying Eq.~\eqref{eq:appendix:state_derivative_intermediate} with the state vector \(\ket{\pPsi_{m \vb k'}} \) and summing over all states yields:
\begin{multline}
	\overbrace{\sum_{m\vb k'} \dyad*{\pPsi_{m\vb k'}}{\pPsi_{m\vb k'}} \paw{S}}^{= \id} \ket{\partial_\tau \pPsi_\nk}
	 =
	\sum_{m\vb k'}' \frac{\ket{\pPsi_{m\vb k'}} \paw{h}_{m\vb k' \nk, \tau}}{\varepsilon_\nk - \varepsilon_{m\vb k'}}
	\\+
	\dyad*{\pPsi_\nk}{\pPsi_\nk} \paw{S} \ket{\partial_\tau \pPsi_\nk}
	\, .
\end{multline}
The last term can be recast in a form that does not involve any derivatives of state vectors.
To show this, one starts from the PAW orthogonality relation, Eq.~\eqref{eq:appendix:orthogonality_constraint}, and takes the first derivative:
\begin{equation}
	\partial_\tau \qty(\braket{\pPsi_\nk | \paw{S} | \pPsi_\nk}) = 0
	\, .
\end{equation}
Then, one applies the chain rule,
\begin{multline}
	\braket{\partial_\tau \pPsi_\nk | \paw{S} | \pPsi_\nk}
	+
	\braket{\pPsi_\nk | \paw{S} | \partial_\tau \pPsi_\nk}
	\\ =
	- \braket{\pPsi_\nk | \partial_\tau \paw{S} | \pPsi_\nk}
	\, ,
\end{multline}
and uses the phase freedom of Bloch orbitals to obtain:
\begin{equation}
	\braket{\pPsi_\nk | \paw{S} | \partial_\tau \pPsi_\nk} = - \frac{1}{2} \paw{h}^\text{S}_{\nk \nk, \tau}
	\, .
\end{equation}
Thus, the derivative of a pseudo orbital can be expressed as:
\begin{equation}
	\label{eq:appendix:state_derivative}
	\ket{\partial_\tau \pPsi_\nk}
	 =
	\sum_{m\vb k'}' \frac{\ket{\pPsi_{m\vb k'}} \paw{h}_{m\vb k' \nk, \tau}}{\varepsilon_\nk - \varepsilon_{m\vb k'}}
	\\ -
	\frac{1}{2} \ket{\pPsi_\nk} \paw{h}^\text{S}_{\nk \nk, \tau}
	\, .
\end{equation}
Finally, using Eq.~\eqref{eq:appendix:KS_first_derivative}, \eqref{eq:appendix:KS_second_derivative_beginning}, and \eqref{eq:appendix:state_derivative}, one finds the second derivative of the KS eigenvalues to be:
\begin{multline}
	\partial_\tau \partial_{\tau'} \varepsilon_\nk
	=
	\braket{\pPsi_\nk | \partial_\tau \partial_{\tau'} \paw{H} - \varepsilon_\nk \partial_\tau \partial_{\tau'} \paw{S} | \pPsi_\nk}
	\\ +
	\sum_{m\vb k'}'
	\frac{
		\paw{h}^*_{m \vb k' \nk, \tau}
		\paw{h}_{m \vb k' \nk, \tau'}
	}
	{\varepsilon_{\nk} - \varepsilon_{m \vb k'}}
	+
	\sum_{m\vb k'}'
	\frac{
		\paw{h}^*_{m \vb k' \nk, \tau'}
		\paw{h}_{m \vb k' \nk, \tau}
	}
	{\varepsilon_{\nk} - \varepsilon_{m \vb k'}}
	\\ -
	\paw{h}_{\nk, \nk, \tau} \paw{h}^\text{S}_{\nk, \nk, \tau'}
	-
	\paw{h}_{\nk, \nk, \tau'} \paw{h}^\text{S}_{\nk, \nk, \tau}
	\, .
\end{multline}
This result is consistent with the split into two contributions in the main text, namely Eq.~\eqref{eq:FM_contribution} and~\eqref{eq:DW_contribution}.

\section{Derivation of Self-energy Expressions} \label{sec:appendix:self_energy}

For brevity's sake, the following derivation is only performed on the simplest term of the FM contribution, Eq.~\eqref{eq:FM_contribution}.
The remaining terms can be derived in a completely analogous way but would require even more space.
The term in question is \(\paw{h}_{\nk \nk, \tau} \paw{h}^\text{S}_{\nk \nk, \tau'} \) and its contribution to the phonon-induced energy shift is:
\begin{equation}
	\Delta \varepsilon^\text{FM,a}_\nk \qty(T)
	\equiv
	\frac{1}{2} \sum_{\tau \tau'} \paw{h}_{\nk \nk, \tau} \paw{h}^\text{S}_{\nk \nk, \tau'} \ev{u_\tau u_{\tau'}}_T
	\, .
\end{equation}
From there, the partial derivatives with respect to individual atomic displacements, appearing in \(\paw{h}_{\nk \nk, \tau} \) and \(\paw{h}^\text{S}_{\nk \nk, \tau'} \), are rewritten using the phonon differential operator, \(\partial_{\nu \vb q} \), defined in Eq.~\eqref{eq:derivative_phonons}.
Equation~\eqref{eq:derivative_ions} allows for a simple substitution of the partial derivatives, resulting in:
\begin{widetext}
	\begin{multline}
		\Delta \varepsilon^\text{FM,a}_\nk \qty(T)
		=
		\frac{1}{\hbar}
		\sum_{l\kappa\alpha} \sum_{l'\kappa'\beta} \sum_{\nu_1 \nu_2}
		\bzint \kint{q_1} \bzint \kint{q_2}
		\sqrt{m_\kappa m_{\kappa'} \omega_{\nu_1 \vb q_1} \omega_{\nu_2 \vb q_2}}
		\eul^{-\ci \vb q_1 \vdot \vb R_{l\kappa}} \eul^{-\ci \vb q_2 \vdot \vb R_{l'\kappa'}}
		e^*_{\kappa\alpha, \nu_1 \vb q_1} e^*_{\kappa'\beta, \nu_2 \vb q_2}
		\\ \cross
		\braket{\pPsi_\nk | \partial_{\nu_1 \vb q_1} \paw{H} - \varepsilon_\nk \partial_{\nu_1 \vb q_1} \paw{S} | \pPsi_\nk}
		\braket{\pPsi_\nk | \partial_{\nu_2 \vb q_2} \paw{S} | \pPsi_\nk}
		\ev{u_{l\kappa\alpha} u_{l'\kappa'\beta}}_T
		\, .
	\end{multline}
Next, the thermal expectation value of the atomic displacements is expressed in phonon modes by using Eq.~\eqref{eq:thermal_expect}, cancelling out the ionic masses.
In addition, it is now possible to evaluate the sums over all cells with indices \(l \) and \(l' \):
	\begin{multline}
		\Delta \varepsilon^\text{FM,a}_\nk \qty(T)
		=
		\sum_{\kappa\alpha} \sum_{\kappa'\beta} \sum_{\nu \nu_1 \nu_2}
		\bzint \kint{q} \bzint \kint{q_1} \bzint \kint{q_2}
		\frac{\sqrt{\omega_{\nu_1 \vb q_1} \omega_{\nu_2 \vb q_2}}}{\omega_{\nu\vb q}}
		\qty(n_{\nu\vb q}\qty(T) - \frac{1}{2})
		e^*_{\kappa\alpha, \nu_1 \vb q_1} e^*_{\kappa'\beta, \nu_2 \vb q_2}
		e_{\kappa\alpha, \nu \vb q} e^*_{\kappa'\beta, \nu \vb q}
		\\ \cross
		\braket{\pPsi_\nk | \partial_{\nu_1 \vb q_1} \paw{H} - \varepsilon_\nk \partial_{\nu_1 \vb q_1} \paw{S} | \pPsi_\nk}
		\braket{\pPsi_\nk | \partial_{\nu_2 \vb q_2} \paw{S} | \pPsi_\nk}
		\underbrace{\sum_l \eul^{\ci \vb R_{l\kappa} \vdot \qty(\vb q - \vb q_1)}}_{=\BZvol \diracv{\vb q - \vb q_1}}
		\underbrace{\sum_{l'} \eul^{-\ci \vb R_{l'\kappa'} \vdot \qty(\vb q + \vb q_2)}}_{=\BZvol \diracv{\vb q + \vb q_2}}
		\, .
	\end{multline}
After integrating out the Dirac delta functions, the sums over atom and Cartesian indices are performed.
Since the phonon eigenvectors can always be chosen to be orthonormal, the result is:
	\begin{multline}
		\label{eq:appendix:FM_derivation_eigenvectors}
		\Delta \varepsilon^\text{FM,a}_\nk \qty(T)
		=
		\sum_{\nu \nu_1 \nu_2}
		\bzint \kint{q}
		\qty(n_{\nu\vb q}\qty(T) - \frac{1}{2})
		\braket{\pPsi_\nk | \partial_{\nu_1 \vb q} \paw{H} - \varepsilon_\nk \partial_{\nu_1 \vb q} \paw{S} | \pPsi_\nk}
		\braket{\pPsi_\nk | \partial_{\nu_2 -\vb q} \paw{S} | \pPsi_\nk}
		\\ \cross
		\underbrace{\sum_{\kappa\alpha} e_{\kappa\alpha, \nu \vb q} e^*_{\kappa\alpha, \nu_1 \vb q}}_{=\delta_{\nu \nu_1}}
		\underbrace{\sum_{\kappa'\beta} e_{\kappa'\beta, \nu_2 \vb q} e^*_{\kappa'\beta, \nu \vb q}}_{=\delta_{\nu \nu_2}}
		\, .
	\end{multline}
\end{widetext}
In the final step, a constraint is put on the phonon wave vector, \(\vb q \).
The Bloch theorem implies that for a matrix element involving some lattice-periodic operator, \(\hat{\mathcal{O}} \), the following relation must hold:
\begin{equation}
	\braket{\Psi_{m\vb k'} | \partial_{\nu\vb q} \hat{\mathcal{O}} | \Psi_\nk}
	\Rightarrow
	\vb k' = \vb k + \vb q
	\, .
\end{equation}
Therefore, since \(\vb k \) is found on both sides of each matrix element, the phonon wave vector must be confined to the Brillouin-zone center:
\begin{equation}
	\Delta \varepsilon^\text{FM,a}_\nk \qty(T)
	=
	\sum_\nu
	\paw{g}_{nn \vb k, \nu \vb 0}
	\paw{g}^\text{S}_{nn \vb k, \nu \vb 0}
	\qty(n_{\nu \vb 0}\qty(T) + \frac{1}{2})
	\, .
\end{equation}

In Eq.~\eqref{eq:FM_contribution}, the term \(\paw{h}_{\nk \nk, \tau'} \paw{h}^\text{S}_{\nk \nk, \tau} \) results in exactly the same energy shift, \(\Delta \varepsilon^\text{FM,a}_\nk \qty(T) \).
As mentioned earlier, the two remaining terms involving sums over all states can be derived analogously.
The only noticeable addition is the fact that \(\partial_{\nu -\vb q} = \partial^*_{\nu\vb q} \), which can be used to rewrite one matrix element as a complex conjugate.
Finally, putting everything together results in Eq.~\eqref{eq:FM_self_energy}.

The DW self-energy is derived by employing the rigid-ion approximation and using the following acoustic sum rule:
\begin{equation}
	\label{eq:acoustic_sum_rule}
	\sum_{l'\kappa'} \pdv{\varepsilon_\nk}{u_{l\kappa\alpha}}{u_{l'\kappa'\beta}}
	= 0
	\, .
\end{equation}
The latter can be used to relate FM and DW contributions with the help of Eq.~\eqref{eq:FM_DW_contribution_split},
\begin{equation}
	\sum_{l'\kappa'} \delta \varepsilon^\text{DW}_{\nk, l\kappa\alpha l'\kappa'\beta}
	=
	- \sum_{l'\kappa'} \delta \varepsilon^\text{FM}_{\nk, l\kappa\alpha l'\kappa'\beta}
	\, ,
\end{equation}
while the former implies that all off-site terms in the DW contribution can be approximated as zero:
\begin{equation}
	\delta \varepsilon^\text{DW}_{\nk, l\kappa\alpha l'\kappa'\beta}
	\approx 0
	\qq{for}
	\qty(l\kappa) \neq \qty(l'\kappa')
	\, .
\end{equation}
This leads to a simplified expression for the DW self energy that only contains first-order derivatives and is suitable for numerical calculations:
\begin{multline}
	\Delta \varepsilon^\text{DW}_\nk\qty(T)
	=
	\frac{1}{2} \sum_{l \kappa \alpha} \sum_{\beta}
	\delta \varepsilon^\text{DW}_{n\vb k, l\kappa\alpha l\kappa\beta}
	\ev{u_{l \kappa \alpha} u_{l \kappa \beta}}_T
	\\ =
	-\frac{1}{2} \sum_{l \kappa \alpha} \sum_{l' \kappa' \beta}
	\delta \varepsilon^\text{FM}_{n\vb k, l\kappa\alpha l'\kappa'\beta}
	\ev{u_{l \kappa \alpha} u_{l \kappa \beta}}_T
	\, .
\end{multline}

Performing the algebraic derivation from this point onwards is analogous to the steps taken in deriving the FM self energy.
A noticeable difference lies in the fact that the ionic displacements are now restricted to the same atom, \(\qty(l\kappa) \).
Just as in the FM case, the derivation limits itself to the simple term \(\paw{h}_{\nk \nk, \tau} \paw{h}^\text{S}_{\nk \nk, \tau'} \) for the sake of brevity.
The corresponding energy shift, \(\Delta \varepsilon^\text{FM,a}_\nk \), then reads:
\begin{widetext}
	\begin{multline}
		\Delta \varepsilon^\text{DW,a}_\nk \qty(T)
		=
		- \sum_{\kappa\alpha} \sum_{\kappa'\beta} \sum_{\nu \nu_1 \nu_2}
		\bzint \kint{q} \bzint \kint{q_1} \bzint \kint{q_2}
		\frac{\sqrt{\omega_{\nu_1 \vb q_1} \omega_{\nu_2 \vb q_2}}}{\omega_{\nu\vb q}}
		\frac{\sqrt{m_\kappa m_{\kappa'}}}{m_\kappa}
		\qty(n_{\nu\vb q}\qty(T) - \frac{1}{2})
		e_{\kappa\alpha, \nu \vb q} e^*_{\kappa\beta, \nu \vb q}
		\\ \cross
		e^*_{\kappa\alpha, \nu_1 \vb q_1} e^*_{\kappa'\beta, \nu_2 \vb q_2}
		\braket{\pPsi_\nk | \partial_{\nu_1 \vb q_1} \paw{H} - \varepsilon_\nk \partial_{\nu_1 \vb q_1} \paw{S} | \pPsi_\nk}
		\braket{\pPsi_\nk | \partial_{\nu_2 \vb q_2} \paw{S} | \pPsi_\nk}
		\underbrace{\sum_l \eul^{-\ci \vb q_1 \vdot \vb R_{l\kappa}}}_{=\BZvol \diracv{\vb q_1}}
		\underbrace{\sum_{l'} \eul^{-\ci \vb q_2 \vdot \vb R_{l'\kappa'}}}_{=\BZvol \diracv{\vb q_2}}
		\, .
	\end{multline}
	Integrating over the Dirac delta functions and rearranging some of the terms yields
	\begin{multline}
		\Delta \varepsilon^\text{DW,a}_\nk \qty(T)
		=
		- \frac{1}{2} \sum_{\kappa\alpha} \sum_{\kappa'\beta} \sum_{\nu}
		\bzint \kint{q}
		\frac{\hbar}{m_\kappa \omega_{\nu\vb q}}
		\qty(n_{\nu\vb q}\qty(T) - \frac{1}{2})
		e_{\kappa\alpha, \nu \vb q} e^*_{\kappa\beta, \nu \vb q}
		\\ \cross
		\underbrace{
			\sum_{\nu_1}
			\sqrt{\frac{2 m_\kappa \omega_{\nu_1\vb 0}}{\hbar}}
			e_{\kappa\alpha, \nu_1 \vb 0}
			\braket{\pPsi_\nk | \partial_{\nu_1 \vb 0} \paw{H} - \varepsilon_\nk \partial_{\nu_1 \vb 0} \paw{S} | \pPsi_\nk}
		}_{\equiv \paw{g}^0_{nn \vb k, \kappa \alpha}}
		\underbrace{
			\sum_{\nu_2}
			\sqrt{\frac{2 m_{\kappa'} \omega_{\nu_2\vb 0}}{\hbar}}
			e_{\kappa'\beta, \nu_2 \vb 0}
			\braket{\pPsi_\nk | \partial_{\nu_2 \vb 0} \paw{S} | \pPsi_\nk}
		}_{\equiv \paw{g}^{\text{S}0}_{nn \vb k, \kappa' \beta}}
		\, ,
	\end{multline}
	where the definitions~\eqref{eq:DW_g0_definition} and~\eqref{eq:DW_gS0_definition} from the main text have been used.
	Finally, one is able to bring the partial energy shift into a form that makes it easier to understand how the DW self energy is obtained:
	\begin{equation}
		\label{eq:appendix:DW_self_energy_almost_final}
		\Delta \varepsilon^\text{DW,a}_\nk \qty(T)
		=
		- \bzint \kint{q} \sum_\nu
		\qty(2 n_{\nu\vb q}\qty(T) + 1)
		\frac{\hbar}{4 \omega_{\nu \vb q}}
		\sum_{\kappa \alpha} \sum_{\kappa' \beta}
		\frac{e_{\kappa\alpha, \nu \vb q} e^*_{\kappa\beta, \nu \vb q}}{m_\kappa}
		\paw{g}^0_{nn\vb k, \kappa \alpha} \paw{g}^{\text{S}0}_{nn\vb k, \kappa' \beta}
		\, .
	\end{equation}
\end{widetext}
The inclusion of the term \(\paw{h}_{\nk \nk, \tau'} \paw{h}^\text{S}_{\nk \nk, \tau} \) simply adds Eq.~\eqref{eq:appendix:DW_self_energy_almost_final} with swapped indices which can be written using \(\Theta_{\kappa \alpha, \kappa' \beta}^{\nu \vb q} \) from Eq.~\eqref{eq:DW_Theta_definition}.
The remaining terms involve a sum over electronic states but are otherwise treated analogously.
In the end, everything accumulates exactly in Eq.~\eqref{eq:DW_self_energy}.

\section{Difference between PAW and AE Electron-phonon Matrix Elements} \label{sec:appendix:ae_vs_paw_epme}

Here, a simple algebraic proof of Eq.~\eqref{eq:ae_vs_paw_epme} is provided.
Starting from the definition of the AE electron-phonon matrix element, the AE orbitals are first expanded using the PAW transformation operators:
\begin{equation} \label{eq:appendix:ae_paw_start}
	\begin{split}
		g_{m \nk,\nu \vb q}
		& =
		\braket{\Psi_{m \vb k + \vb q} | \partial_{\nu \vb q} \hat{H} | \Psi_\nk}
		\\ & =
		\braket{\pPsi_{m \vb k + \vb q} | \pawT^\dagger \partial_{\nu \vb q} \hat{H} \pawT | \pPsi_\nk}
		\, .
	\end{split}
\end{equation}
The product rule of differentiation allows rewriting the operator inside the braket as:
\begin{equation}
	\pawT^\dagger \partial_{\nu \vb q} \hat{H} \pawT
	=
	\partial_{\nu \vb q} \underbrace{\qty(\pawT^\dagger \hat{H} \pawT)}_{\paw{H}}
	- \partial_{\nu \vb q} \pawT^\dagger \hat{H} \pawT
	- \pawT^\dagger \hat{H} \partial_{\nu \vb q} \pawT
	\, .
\end{equation}
Reinsertion of this expression into Eq.~\eqref{eq:appendix:ae_paw_start} yields:
\begin{equation}
	\begin{split}
		g_{m \nk,\nu \vb q}
		& =
		\braket{\pPsi_{m \vb k + \vb q} | \partial_{\nu \vb q} \paw{H} | \pPsi_\nk}
		\\ & -
		\bra{\pPsi_{m \vb k + \vb q}} \partial_{\nu \vb q} \pawT^\dagger
		\underbrace{\hat{H} \pawT \ket{\pPsi_\nk}}_{\varepsilon_\nk \pawT \ket{\pPsi_\nk}}
		\\ & -
		\underbrace{\bra{\pPsi_{m \vb k + \vb q}} \pawT^\dagger \hat{H}}_{\mathclap{\varepsilon_{m \vb k + \vb q} \bra{\pPsi_{m \vb k + \vb q}} \pawT^\dagger}}
		\partial_{\nu \vb q} \pawT \ket{\pPsi_\nk}
		\, ,
	\end{split}
\end{equation}
where the KS equations have been used to substitute \(\hat{H} \) by the eigenvalues.
Finally, the relation
\begin{equation}
	\partial_{\nu \vb q} \paw{S} = \partial_{\nu \vb q} \qty(\pawT^\dagger \pawT) = \partial_{\nu \vb q} \pawT^\dagger \pawT + \pawT^\dagger \partial_{\nu \vb q} \pawT
\end{equation}
is used to rewrite the term containing \(\partial_{\nu \vb q} \pawT^\dagger \pawT \) in terms of the PAW overlap operator and the remaining contribution,
\begin{multline}
	g_{mn\vb k,\nu\vb q}
	=
	\overbrace{\braket{\pPsi_{m \vb k + \vb q} | \partial_{\nu \vb q} \paw{H} - \varepsilon_\nk \partial_{\nu \vb q} \paw{S} | \pPsi_\nk}}^{\paw{g}_{mn\vb k, \nu \vb q}}
	\\ +
	\qty(\varepsilon_\nk - \varepsilon_{m\vb k + \vb q}) \braket{\pPsi_{m\vb k + \vb q} | \pawT^\dagger \partial_{\nu\vb q} \pawT | \pPsi_\nk}
	\, ,
\end{multline}
which concludes the proof.
\newpage
\bibliography{References.bib}

\end{document}